\documentclass[a4paper,11pt]{article}
\pdfoutput=1
\usepackage{graphicx}
\usepackage{jheppub}
\usepackage{orcidlink}
\usepackage{amsmath,amssymb,amsfonts,mathtools,bbold}
\usepackage{braket}
\usepackage[usenames,dvipsnames]{xcolor}
\usepackage{bm}
\usepackage{tikz}
\usepackage{tikz-3dplot}
\usetikzlibrary{arrows.meta, angles, quotes, calc}

\newcommand{\dd}{{\mathrm{d}}}
\newcommand{\LUV}{\Lambda_{\text{UV}}}
\newcommand{\LIR}{\Lambda_{\text{IR}}}

\renewcommand{\Re}{{\operatorname{Re}}}
\newcommand{\Tr}{{\operatorname{Tr}}}

\newcommand{\bsk}{{\boldsymbol{k}}}
\newcommand{\bsp}{{\boldsymbol{p}}}
\newcommand{\bsl}{{\boldsymbol{l}}}
\newcommand{\bsx}{{\boldsymbol{x}}}

\newcommand{\pL}{{p_{\Lambda}}}
\newcommand{\bpL}{{\bar{p}_{\Lambda}}}

\newcommand{\AxisRotator}[1][rotate=0]{
    \tikz [x=0.25cm,y=0.60cm,line width=.2ex,-stealth,#1] \draw (0,0) arc (-150:150:1 and 1);
}

\preprint{UT-Komaba/25-7}

\title{Perturbative unitarity bounds on field-space curvature in de Sitter spacetime: purity vs scattering amplitude}

\author[a]{Qianhang Cai\,\orcidlink{0009-0000-5018-0343},}
\emailAdd{qcai@g.ecc.u-tokyo.ac.jp}

\author[a]{Tomoya Inada\,\orcidlink{0009-0009-4444-9110},}
\emailAdd{inada-tomoya@g.ecc.u-tokyo.ac.jp}

\author[a]{Masataka Ishikawa\,\orcidlink{0009-0001-9919-5084},}
\emailAdd{marlborough@g.ecc.u-tokyo.ac.jp}

\author[a,b]{Kanji Nishii\,\orcidlink{0009-0008-2153-0365}}
\emailAdd{nishii-kanji743@g.ecc.u-tokyo.ac.jp}

\author[a]{and Toshifumi Noumi\,\orcidlink{0000-0003-0628-8554}}
\emailAdd{tnoumi@g.ecc.u-tokyo.ac.jp}

\affiliation[a]{Graduate School of Arts and Sciences, The University of Tokyo\\
Komaba, Meguro-ku, Tokyo 153-8902, Japan}
\affiliation[b]{Quantum Computing Center, Keio University\\
3-14-1 Hiyoshi, Kohoku-ku, Yokohama, Kanagawa, 223-8522, Japan}

\abstract{
We study perturbative unitarity bounds on the field-space curvature in de Sitter spacetime, using the momentum-space entanglement approach recently proposed by Duaso Pueyo, Goodhew, McCulloch, and Pajer. As an illustration, we perform a perturbative computation of the purity in two-scalar models and compare the resulting unitarity bounds with those obtained via a flat space approximation. In particular, we find that perturbative unitarity imposes an upper bound on the field-space curvature of the Hubble scale order, in addition to a bound analogous to the flat space result. This reflects the thermal nature of de Sitter spacetime. We also discuss generalizations to higher-dimensional field spaces.
}

\begin{document} 
\setcounter{tocdepth}{2}
\maketitle
\flushbottom

\section{Introduction}

Effective field theory (EFT) provides a universal framework for describing physical phenomena at energy scales of interest, without requiring detailed knowledge of the underlying high-energy dynamics. By identifying the relevant symmetries and dynamical degrees of freedom, one can systematically construct an effective Lagrangian that captures the low-energy dynamics.

Notably, EFT becomes more powerful when combined with the concept of ultraviolet (UV) completion and fundamental principles such as unitarity. In particular, the S-matrix unitarity offers a powerful criterion for quantifying the UV cutoff scale and searching for new physics required for UV completion. Historically, the Higgs boson was predicted to restore unitarity in the high-energy scattering of weak bosons~\cite{lee1991weak,Lee:1977yc,Dicus:1973gbw, Chanowitz:1985hj}. Similarly, string theory, as a UV-complete theory of gravity, emerged from S-matrix theory~\cite{Veneziano:1968yb}. Moreover, the study of UV completion has in turn revealed that not every EFT is UV completable, leading to nontrivial UV constraints on low-energy effective theories (see, e.g.,~\cite{deRham:2022hpx} for a review article). 

While the S-matrix unitarity offers a fundamental tool to theoretically connect various scales in nature, cosmology is beyond its scope because scattering amplitudes are not well-defined in cosmological backgrounds. To address this limitation, several approaches have been taken so far: A pragmatic approach is to employ the flat space approximation and directly apply implications of the S-matrix unitarity to cosmological models~\cite{Baumann:2015nta, Kim:2019wjo, Grall_2020, Melville_2020, Kim:2021pbr, Freytsis:2022aho, Grall:2021xxm}. While this approach has limitations in applicability, the approximation can be justified at least physically as long as our focus is, e.g., on effective couplings generated by the UV dynamics well above the Hubble scale or, in other words, by the dynamics well inside the horizon.

A more challenging direction would be to set up the bootstrap program in cosmology, motivated by recent progress in the S-matrix bootstrap and the conformal bootstrap. Aiming at this ultimate goal, unitarity and (non-)analyticity of cosmological correlators and wavefunctions of the universe have been studied intensively under the slogan of the Cosmological Bootstrap~\cite{Arkani-Hamed:2018kmz,Baumann:2019oyu,Baumann:2020dch,Arkani-Hamed:2017fdk,Benincasa:2018ssx,Sleight:2019mgd,Sleight:2019hfp,Goodhew:2020hob,Cespedes:2020xqq,Pajer:2020wxk,Jazayeri:2021fvk,Bonifacio:2021azc,Melville:2021lst,Goodhew:2021oqg,Pimentel:2022fsc,Jazayeri:2022kjy,Qin:2022fbv,Xianyu:2022jwk,Wang:2022eop,Qin:2023ejc,Stefanyszyn:2023qov,DuasoPueyo:2023kyh,Cespedes:2023aal,Bzowski:2023nef,Arkani-Hamed:2023kig,Grimm:2024mbw,Aoki:2024uyi,Stefanyszyn:2024msm,Liu:2024xyi,Goodhew:2024eup,Ghosh:2024aqd,Lee:2024sks,Cespedes:2025dnq,Pimentel:2025rds,Stefanyszyn:2025yhq,Qin:2025xct}. There are also alternative attempts to define the notion of the S-matrix itself to cosmological backgrounds~\cite{Mack:2009mi, Penedones:2010ue, Marolf:2012kh, Melville:2023kgd, Donath:2024utn, Melville:2024ove, Taylor:2024vdc, Ferrero:2021lhd, Mandal:2019bdu, Spradlin:2001nb, Mei:2024sqz}. These developments motivate further studies toward refinement of the S-matrix unitarity as a guiding principle in cosmology.

Building upon this insight, an interesting approach was proposed recently in~\cite{Pueyo:2024twm} to utilize entanglement measures such as purity and entanglement entropy to derive unitarity constraints on cosmological models (see also~\cite{Colas:2022kfu, Colas:2024xjy, Burgess:2024eng,Ueda:2024cyf, Balasubramanian:2011wt, Aoude:2024xpx, Cheung:2023hkq,Peschanski:2016hgk,Kowalska:2024kbs,Brahma:2023lqm,Boutivas:2023mfg} for related developments). A key of this approach is in the fact that interactions induce entanglement between momentum modes, leading to an analogy between entanglement measures and scattering amplitudes. Crucially, this framework is applicable even in curved spacetime as long as the density matrix is well-defined. Based on this approach, perturbative unitarity bounds in inflationary backgrounds were studied in particular.

In this paper, we apply the momentum-space entanglement approach of~\cite{Pueyo:2024twm} to derive perturbative unitarity bounds on field-space curvature of nonlinear sigma models, which widely appear for example as effective theories of (pseudo-)Nambu-Goldstone bosons, on a fixed de Sitter background. Unlike the original paper, we study the unitarity bounds on purity without taking the superhorizon limit, which allows us to perform detailed analysis of the bounds that interpolate the flat space analysis and the superhorizon analysis. Interestingly, in addition to a bound similar to the flat space result, we find that the perturbative unitarity gives an upper bound on the field-space curvature of order $H^{-2}$ in terms of the Hubble parameter $H$, which reflects the thermal nature of de Sitter spacetime.

\paragraph{Outline:} 

This paper is organized as follows:
In Sec.~\ref{review_purity}, we review the momentum-space entanglement approach to perturbative unitarity proposed in~\cite{Pueyo:2024twm}. In particular, we introduce a perturbative formula of purity and its unitarity condition.
In Sec.~\ref{sec:flat}, we study the perturbative unitarity bounds in flat spacetime and show that the UV cutoff is set by the field-space curvature, similarly to the bounds obtained from scattering amplitudes.
In Sec.~\ref{sec:deSitter}, we extend the analysis to de Sitter spacetime. In addition to a bound similar to the flat space result, we find an upper bound on the field-space curvature at the order of the Hubble scale.
We conclude our analysis in Sec.~\ref{sec:conclusions}.
Technical details are collected in the Appendices.

\paragraph{Convention:}
Throughout the paper, we adopt the metric signature $(-, +, +, +)$, Greek letters $\mu, \nu, \cdots$ to denote spacetime indices, and shorthand notations for the integration measure,
\begin{align}
    \int_{\bsk}= \int\frac{\dd^3 \bsk}{(2\pi)^3}\,,
\qquad    
\int_{\bsk_1\cdots\bsk_n}=\prod_{i=1}^n \int\frac{\dd^3 \bsk_i}{(2\pi)^3}\,.
\end{align}
We work in natural units, setting $c = \hbar = 1$.

\section{Perturbative unitarity bounds from purity: a brief review}\label{review_purity}

This section gives a brief review of the momentum-space entanglement approach to perturbative unitarity bounds proposed in~\cite{Pueyo:2024twm}. In particular, we consider purity that quantifies the entanglement between a system of interest and its complement (environment). In Sec.~\ref{purity_QFT}, we first introduce the concept of purity in quantum field theory (QFT) and briefly explain how it can be evaluated using the wavefunction representation. Then, in Sec.~\ref{purity_non-linear}, we consider multi-scalar models with nonzero field-space curvature and provide a concrete formula for the purity, which is used in the following sections.

\subsection{Momentum-space entanglement and purity in EFT}\label{purity_QFT}

\paragraph{Purity.}

Consider a quantum system in a pure state represented by a density matrix\footnote{
In \cite{Pueyo:2024twm}, the density matrix was defined without being canonically normalized to carefully discuss regularization for the continuum and infinite-volume limit. The density matrix here should also be understood under the same regularization, even though we assume canonical normalization for visual clarity.},
\begin{align}
\rho=\ket{\Omega}\bra{\Omega}\,,
\quad
\Tr \, \rho = 1\,.
\end{align}
If we split the Hilbert space into a system $\mathcal{S}$ and its complement (environment) $\mathcal{E}$, the reduced density matrix $\rho_{\mathrm{R}}$ after tracing out the environment sector $\mathcal{E}$ is defined by
\begin{align}
\rho_{\mathrm{R}} \coloneqq \Tr_{\mathcal{E}}\,\rho\,,
\quad
\Tr_{\mathcal{S}}\,\rho_{\mathrm{R}}=1\,,
\end{align}
where $\Tr_{\mathcal{E}}$ and $\Tr_{\mathcal{S}}$ denote the trace over the environment sector and that of the system sector, respectively. The entanglement between the system and the environment can be quantified by the purity $\gamma$ defined by
\begin{align}
    \gamma \coloneqq \Tr_{\mathcal{S}}\,\rho_{\mathrm{R}}^2\,.
\end{align}
The norm positivity, a requirement of unitarity, implies that $0\leq\gamma\leq1$, which will be used in the following discussion as a consistency requirement of the theory. Also, the upper bound is saturated at $\gamma=1$ if and only if $\rho_{\mathrm{R}}$ is a pure state, hence it is called purity.

\paragraph{Momentum-space entanglement.}

Next, we consider momentum-space entanglement in QFT. For this, it is convenient to employ the field eigenstate $\ket{\phi}$ as a basis of the Hilbert space and express the density matrix $\rho$ at a given time $\eta=\eta_*$ as
\begin{align}
    \rho 
    & = \ket{\Omega} \bra{\Omega}\notag\\    
    & =\int \mathcal{D} \phi \mathcal{D} \phi' 
    \ket{\phi} \braket{\phi|\Omega} \braket{\Omega|\phi'} \bra{\phi'}
    \notag\\
    & = \int \mathcal{D} \phi \mathcal{D} 
    \phi' \rho_{\phi \phi'}\ket{\phi}\bra{\phi'}
    \,,
\end{align}
where we define the components of the density matrix as $\rho_{\phi \phi'} \coloneqq \braket{\phi|\Omega}\braket{\Omega|\phi'}$. Here and in the rest of this subsection, we focus on a single real scalar model for illustration, but its extension to general setups is straightforward.

For the concrete analysis of momentum-space entanglement, we choose the Fourier modes $\phi_{ \bsp}$ and $\phi_{- \bsp}$ with $\bsp\neq 0$ as the system $\mathcal{S}$. The corresponding environment $\mathcal{E}$ consists of all the remaining Fourier modes $\phi_{\bsk}$ with $\bsk\neq\pm \bsp$.
Then, the reduced density matrix $\rho_{\mathrm{R}}(\bsp)$ reads
\begin{align}
\label{rho_R_QFT}
    \left(\rho_{\mathrm{R}}(\bsp)\right)_{\phi \phi'} =\left( \prod_{\bsk\in \mathcal{E}}\int \dd\phi_{\bsk}\right)
    \rho_{\phi\phi'}\big|_{\phi_{\bsk}=\phi'_{\bsk}}\,,
\end{align}
where the index $\phi$ of the reduced density matrix $\left(\rho_{\mathrm{R}}(\bsp)\right)_{\phi \phi'}$ denotes the system modes $\phi_{\bsp},\, \phi_{-\bsp}$ collectively, and similarly for $\phi'$. The purity $\gamma(\bsp)$ is now given by
\begin{align}
\label{gamma_QFT}
    \gamma(\bsp) = \int \dd\phi_{\bsp}\,\dd\phi_{-\bsp} \,\dd\phi'_{\bsp}\,\dd\phi'_{-\bsp} \,\, (\rho_{\mathrm{R}})_{\phi \phi'} \, (\rho_{\mathrm{R}})_{\phi'\phi}\,.
\end{align}

\paragraph{Application to EFT.}

The purity defined in this manner can be used to study the validity of effective field theories (EFTs).
Suppose that the EFT has a UV cutoff $\LUV$ and an IR cutoff $\LIR$. Then, all the Fourier modes $\phi_{\bsk}$ (both system and environment) have to reside in the range,
\begin{align}
    \LIR\leq E(\bsk)\leq \LUV\,,
\end{align}
where $E(\bsk)$ is the energy associated to the Fourier mode $\bsk$. To make the cutoff-dependence manifest, let us denote the purity by $\gamma(\LIR,\LUV,\bsp)$. In this language, the unitarity constraints are given by
\begin{align}
\label{purity_bound}
0\leq \gamma(\LIR,\LUV,\bsp)\leq1
\qquad
\text{for all}
\quad
\bsp
\quad
\text{with}
\quad
E(\bsp)\in[\LIR,\LUV]\,.
\end{align}
Its violation signals breakdown of the EFT, hence we can use the purity bound~\eqref{purity_bound} to identify the maximum energy range of validity of the EFT\footnote{
Alternatively, if the energy scale of interest is specified and the cutoff scales $\LIR$ and $\LUV$ are given, one can interpret the bounds~\eqref{purity_bound} as consistency conditions on the model parameters such as the particle spectrum and the coupling constants.
}.

\paragraph{Wavefunction representation.}

In practical computations of the purity, it is convenient to introduce the Schrödinger wave functional $\Psi[\phi]$, which is defined by the inner product of the state $\ket{\Omega}$ and the field eigenstate $\ket{\phi}$ at the time $\eta=\eta_*$ as
\begin{align}
    \Psi[\phi] \coloneqq \braket{\phi|\Omega} \,.
\end{align}
In this language, the density matrix reads
\begin{align}
\label{rho_wave}
\rho_{\phi\phi'}=\Psi[\phi]\left(\Psi[\phi']\right)^*\,.
\end{align}
For perturbative computations of the purity, we expand the wavefunction in the Fourier space (of the spatial coordinates) as
\begin{align}
     \Psi[\phi]\propto\exp\left[
    -\sum_{n=2}^{\infty}\frac{1}{n!}\int_{\bsk_1, \cdots, \bsk_n}(2\pi)^3\delta^{3}\Bigg(\sum_{j=1}^{n}\bsk_j\Bigg)\psi_n(\bsk_1, \cdots, \bsk_n)\phi_{\bsk_1}\cdots\phi_{\bsk_n}
    \right] \label{wavefunction}\,,
\end{align}
where $\phi_{\bsk}$ is the Fourier mode of $\phi$ and the kernels $\psi_n(\bsk_1, \cdots \bsk_n)$ are called wavefunction coefficients. We further assume translation invariance along the spatial directions of the theory and the state $\ket{\Omega}$. Besides, an overall constant factor has been suppressed, as it does not affect the subsequent analysis.

\subsection{Perturbative formula for purity}
\label{purity_non-linear}

In this subsection, we provide a concrete formula for the perturbative computation of purity in scalar EFTs with nonzero field-space curvature.  Throughout the paper, we focus on the tree-level analysis and discuss implications of perturbative unitarity.

\paragraph{EFT setup.}

Consider an EFT of $N$ real scalar fields $\phi^I\,(I=1, 2,\cdots,N)$ with the following effective action:
\begin{align}
    S 
    & = \int \dd^4x \sqrt{-g} 
    \left[
    -\frac{1}{2}\sum_{I,J} G_{IJ}(\phi) \,
    \partial_{\mu} \phi^I \partial^{\mu} \phi^J -V(\phi)
    \right]\,,
\end{align}
where $G_{IJ}(\phi)$ is the field-space metric and $V(\phi)$ is the potential. If we choose locally flat coordinates of the field space, the field-space metric can be expanded as
\begin{align}
    G_{IJ}(\phi) = \delta_{IJ} - \sum_{K,L}  C_{IJKL} \phi^K \phi^L + \mathcal{O}(\phi^3)\,,
\end{align}
where $C_{IJKL}$ are constants. Since our main focus is on the field-space curvature, we choose a simple quadratic potential and study the following model:
\begin{align}
    S 
    & = \int\dd^4x \sqrt{-g}
    \left[-\frac{1}{2}\sum_I \Big((\partial_{\mu}\phi^I)^2 + m_I^2 (\phi^I)^2\Big) + \frac{1}{2} \sum_{I,J,K,L} C_{IJKL} \phi^K \phi^L \partial_{\mu} \phi^I \partial^{\mu} \phi^J  +\cdots \right] 
    \,,\label{nonlinear_S}
\end{align}
where the dots stand for higher order terms in $\phi^I$ and they are irrelevant in the following analysis. We study this model in homogeneous and isotropic spacetime,
\begin{align}\label{hom_iso}
    \dd s^2 = a(\eta)^2(-\dd \eta^2 + \dd \bsx^2)\,.
\end{align}
More specifically, we consider flat spacetime in Sec.~\ref{sec:flat} and de Sitter spacetime in Sec.~\ref{sec:deSitter}. As for the state $\ket{\Omega}$, we consider the free theory vacuum for flat spacetime and the Bunch-Davies vacuum for de Sitter spacetime, respectively.

\paragraph{Wavefunction.}

In this model, the tree-level wavefunction takes the form,
\begin{align}
\Psi[\phi]    
& \propto \exp\Bigg[
    -\frac{1}{2}\int_{\bsk_1, \bsk_2}(2\pi)^3\delta^{3}(\bsk_1+\bsk_2)\sum_{I,J}\psi_{IJ}(\bsk_1,\bsk_2)\phi^I_{\bsk_1} \phi^J_{\bsk_2}
    \notag
    \\
    \notag
    &\qquad\quad\,\,\,\,
    - \frac{1}{4!} \int_{\bsk_1, \cdots, \bsk_4}
    (2\pi)^3\delta^{3}\Bigg(\sum_{j=1}^{4}\bsk_j\Bigg)
    \sum_{I,J,K,L} \psi_{IJKL}(\bsk_1,\bsk_2,\bsk_3,\bsk_4)\phi^I_{\bsk_1} \phi^J_{\bsk_2} \phi^K_{\bsk_3} \phi^L_{\bsk_4} 
    \\
    &\qquad\quad\,\,\,\,
    +\mathcal{O}(\phi^5)\Bigg]
   \,,
\end{align}
where $\psi_{IJ}$ and $\psi_{IJKL}$ are defined symmetrically in the field indices $I,J,K,L$ and momenta $\bsk_i$. From the standard perturbation theory, the wavefunction coefficients at $\eta=\eta_*$ are given in terms of the bulk-to-boundary propagator $K_{\phi^I}(k;\eta)$, whose concrete form is shown later, as
\begin{align}
\label{2pt_wave}
    \psi_{IJ}(\bsk,-\bsk) & = \delta_{IJ}\,\psi_{II}(k)
    \quad
    \text{with}
    \quad
    \psi_{II}(k)=
    -i a(\eta)^2 \partial_\eta \log K_{\phi^I}(k;\eta) \Big|_{\eta=\eta_*}
\end{align}
and
\begin{align}\notag
    & \psi_{IJKL}(\bsk_1,\bsk_2,\bsk_3,\bsk_4)\\
    &= \frac{i}{2}C_{IJKL}\int_{-\infty}^{\eta_*}\dd \eta\, a(\eta)^2
    \Big[ K_{\phi^K}(k_3;\eta) K_{\phi^L}(k_4;\eta) \notag\\
    & \hspace{60pt} \times \left(\partial_{\eta} K_{\phi^I}(k_1;\eta) \partial_{\eta} K_{\phi^J}(k_2;\eta) + (\bsk_1\cdot\bsk_2) K_{\phi^I}(k_1;\eta) K_{\phi^J}(k_2;\eta)\right)\Big]
\nonumber
\\
    &\quad
    +\text{ ($23$ perm.)}
    \,,\label{psi4_general}
\end{align}
where the last line denotes $23$ terms obtained by permutations of the field index-momentum pairs $(I,\bsk_1)$, $(J,\bsk_2)$, $(K,\bsk_3)$, $(L,\bsk_4)$ that are necessary to symmetrize $\psi_{IJKL}$.

\paragraph{Purity and perturbative unitarity bound.}

Finally, we provide a formula for the purity. Let us choose the Fourier modes $\phi^{\bar{I}}_{\bsp}$ and $\phi^{\bar{I}}_{-\bsp}$ of the species label $\bar{I}$ as the system of interest. Then, from the definition~\eqref{rho_R_QFT}-\eqref{gamma_QFT} and the wavefunction representation of the density matrix~\eqref{rho_wave}, the purity can be evaluated at the tree level as\footnote{
See the original paper~\cite{Pueyo:2024twm} for details of the diagrammatic method for the purity computation.
}
\begin{align}
     \gamma_{\phi^{\bar{I}}}&(\LIR, \LUV, \bsp)=1-I_{\phi^{\bar{I}}}(\LIR, \LUV, \bsp)
\end{align}
with $I_{\phi^{\bar{I}}}(\LIR, \LUV, \bsp)$ given by\footnote{
$\phi^{\bar{I}}_{\pm\bsp}$ are in the system sector, so that the momentum integral over the environment modes of $\phi^I$ has to be performed such that $\bsk_i\neq\bsp$ to be precise. However, this gives a measure-zero effect and negligible, so that we do not care in the following analysis.
}
\begin{align}
    I_{\phi^{\bar{I}}}(\LIR, \LUV, \bsp)
    &= \frac{1}{24} \int_{\bsk_1, \bsk_2} 
   \sum_{J,K,L} 
   \frac{\left| \psi_{\bar{I}JKL}(\bsp, \bsk_1, \bsk_2,\bsk_3) \right|^2}
   {\Re \,\psi_{\bar{I}\bar{I}}(p) \,\Re\, \psi_{JJ}(k_1) \,\Re \,\psi_{KK}(k_2) \,\Re \,\psi_{LL}(k_3)}\,,
   \label{4ptpurity}
\end{align}
where $\bsk_3=-(\bsp+\bsk_1+\bsk_2)$ and the integral regions of $\bsk_{1,2}$ are defined such that
\begin{align}
\LIR\leq E(\bsp),E(\bsk_1),E(\bsk_2),E(\bsk_3)\leq\LUV
\,.
\end{align}
The unitarity bound on the purity $0 \leq \gamma_{\phi^{\bar{I}}} \leq 1$ implies 
\begin{align}
0\leq I_{\phi^{\bar{I}}}(\LIR, \LUV, \bsp)\leq 1\,.
\end{align}
As we see shortly, $\Re \,\psi_{II}>0$, so that $I_{\phi^{\bar{I}}}>0$ is trivially satisfied. However, $I_{\phi^{\bar{I}}} \leq 1$ gives a non-trivial bound that can be used to derive the perturbative unitarity bound on the coupling constant $C_{IJKL}$ and the cutoff scale.

\section{Perturbative unitarity bounds in flat spacetime} \label{sec:flat}

We begin with the flat spacetime, for which perturbative unitarity bounds on the field-space curvature were already studied using scattering amplitudes (see, e.g.,~\cite{Nagai:2019tgi}). To be concrete, this section focuses on the following two-scalar model as a simple case of \eqref{nonlinear_S}:
\begin{align}
S=\int \dd^4x\left[
-\frac{1}{2}(\partial_\mu \phi)^2
-\frac{1}{2}m_\phi^2\phi^2
-\frac{1}{2}(\partial_\mu\sigma)^2
-\frac{1}{2}m_\sigma^2\sigma^2
+\frac{\epsilon}{4f^2} \sigma^2(\partial_\mu\phi)^2
\right]\,,\label{model_flat}
\end{align}
where $\epsilon=\pm1$ is the sign of the field-space curvature and $f$ is the radius of curvature. The corresponding nonzero component of the coupling constants $C_{IJKL}$ in \eqref{nonlinear_S} reads
\begin{align}
C_{\phi\phi\sigma\sigma} = \frac{\epsilon}{2f^2}\,.
\end{align}
It is well known that the perturbative unitarity bound on scattering amplitudes implies $\LUV\lesssim f$. Below, we reproduce it from the perturbative unitarity bound on the purity.

\subsection{Wavefunction coefficients}

Let us first compute the wavefunction, which takes the form,
\begin{align}
\label{wavefunc_two-scalar}
    \Psi[\phi,\sigma]
    &\propto \exp\Bigg[
    -\frac{1}{2}\int_{\bsk} \left(\psi_{\phi\phi}(k)\phi_{\bsk} \phi_{-\bsk} + \psi_{\sigma\sigma}(k)\sigma_{\bsk} \sigma_{-\bsk}\right)
     \\*
     \notag
    & \qquad\quad\,\,\,\,
    -\frac{1}{4} \int_{\bsk_1, \cdots, \bsk_4}(2\pi)^3\delta^{3}\Bigg(\sum_{j=1}^{4}\bsk_j\Bigg)\psi_{\phi\phi\sigma\sigma}(\bsk_1,\bsk_2,\bsk_3,\bsk_4)\phi_{\bsk_1} \phi_{\bsk_2} \sigma_{\bsk_3} \sigma_{\bsk_4} 
    +\cdots
    \Bigg]\,.
\end{align}
Without loss of generality, we evaluate it at the time $\eta_*=0$. On flat spacetime, the bulk-to-boundary propagators for the Dirichlet problem are simply
\begin{align}
K_{\phi}(k; 
\eta)=e^{iE^{\phi}(\bsk)\,\eta}, \quad 
K_{\sigma}(k; \eta)=e^{iE^{\sigma}(\bsk)\,\eta}
\end{align}
with $E^{\phi}(\bsk)$ and $E^{\sigma}(\bsk)$ given by
\begin{align}
     E^{\phi}(\bsk)=\sqrt{k^2 + m_\phi^2}\,,\quad E^{\sigma}(\bsk)=\sqrt{k^2 + m_\sigma^2}\,.
\end{align}
By applying the formula \eqref{2pt_wave}, the two-point wavefunction coefficients read
\begin{align}
\label{2pt_flat}
    \psi_{\phi\phi}(k)= E^{\phi}(\bsk)\,,\quad
    \psi_{\sigma\sigma}(k)=E^{\sigma}(\bsk)\,.
\end{align}
Similarly, from the formula~\eqref{psi4_general}, the four-point wavefunction coefficient reads
\begin{align}
&\psi_{\phi\phi\sigma\sigma}(\bsk_1,\bsk_2,\bsk_3,\bsk_4) 
 \notag\\
&= \frac{i \epsilon}{f^2} \int_{-\infty}^0 \dd \eta\,
K_{\sigma}(k_3;\eta) K_{\sigma}(k_4;\eta) \bigg[\partial_\eta K_{\phi}(k_1; \eta) \partial_\eta K_{\phi}(k_2; \eta) + (\bsk_1 \cdot \bsk_2) K_{\phi}(k_1;\eta) K_{\phi}(k_2;\eta)\bigg]
\notag
\\
&=-\frac{\epsilon}{f^2}\frac{ E^{\phi}(\bsk_1) E^{\phi}(\bsk_2)-\bsk_1\cdot\bsk_2 }{ E^{\phi}(\bsk_1) + E^{\phi} (\bsk_2)+ E^{\sigma}(\bsk_3) + E^{\sigma} (\bsk_4)}\,.
\end{align}

\subsection{Purity and UV cutoff}
\label{subsec:purity_UV_flat}

We then compute the purity and discuss implications of perturbative unitarity bounds. 

\subsubsection{Bounds on $\gamma_{\phi}$}

First, we consider $\gamma_{\phi}$, choosing the Fourier modes $\phi_\bsp$ and $\phi_{-\bsp}$ as the system. From the formula \eqref{4ptpurity}, the corresponding purity reads\footnote{
Thanks to the rotational symmetry, the momentum-dependence of the purity in our setup is only through the amplitude $p=|\bsp|$ of the momentum $\bsp$ of the system modes. Also, it is convenient for visual clarity to suppress its cutoff-dependence. Hence, we employ the notation $\gamma_{\phi^{\bar{I}}}(p)$ and $I_{\phi^{\bar{I}}}(p)$ for $\gamma_{\phi^{\bar{I}}}(\LIR,\LUV,\bsp)$ and $I_{\phi^{\bar{I}}}(\LIR,\LUV,\bsp)$ in the rest of the paper. 
}
\begin{align}
\gamma_{\phi}(p)  
= 1-I_\phi(p)
\end{align}
with $I_\phi(p)$ given by
\begin{align}
&I_\phi(p)  
= \frac{1}{8} \int_{\bsk_1, \bsk_2}  
   \frac{\left| \psi_{\phi\phi\sigma\sigma}(\bsp, \bsk_1, \bsk_2,\bsk_3) \right|^2}
   {\Re \, \psi_{\phi\phi}(p)\, \Re \, \psi_{\phi\phi}(k_1) \,\Re \, \psi_{\sigma\sigma}(k_2) \,\Re \, \psi_{\sigma\sigma}(k_3)}
   \label{purityphiflat}
   \\
   \notag
    & = \frac{1}{8f^4 E^{\phi}(\bsp)}\int_{\bsk_1, \bsk_2}\frac{1}{
    E^{\phi}(\bsk_1)
    E^{\sigma}(\bsk_2)
    E^{\sigma}(\bsk_3)} \left|\frac{E^{\phi}(\bsp)E^{\phi}(\bsk_1) -\bsp\cdot\bsk_1}{E^{\phi}(\bsp) + E^{\phi}(\bsk_1) + E^{\sigma}(\bsk_2) +E^{\sigma}(\bsk_3)}\right|^2\,,
\end{align}
where $\bsk_3=-(\bsp+\bsk_1+\bsk_2)$ and the integral region of $\bsk_{1,2}$ is defined such that
\begin{align}
\LIR\leq E^\phi(\bsp),E^\phi(\bsk_1),E^\sigma(\bsk_2),E^\sigma(\bsk_3)\leq\LUV
\,.
\end{align}
The integral~\eqref{purityphiflat} is IR-finite for any (non-tachyonic) masses $m_\phi$ and $m_\sigma$, so that we set $\LIR=0$ in the following.

\begin{figure}[t]
    \centering
    \includegraphics[height=7cm]{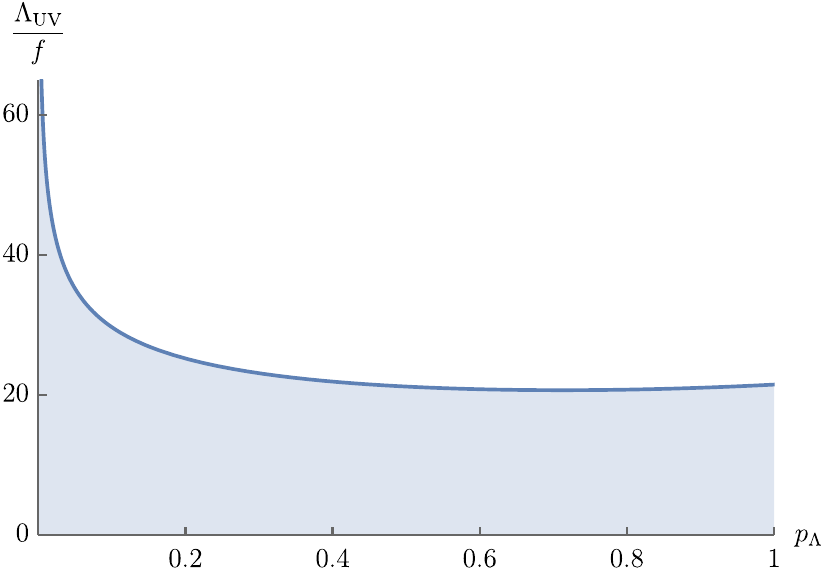}
    \caption{
    The allowed region for $\LUV/ f$ as a function of $\pL$, shown as the shaded area, is derived from the unitarity bound imposed by the purity $\gamma_{\phi}$ in flat spacetime.}
    \label{fig:UVp_phi_flat}
\end{figure}

Due to the presence of the UV cutoff, it is somewhat complicated to perform the integral~\eqref{purityphiflat} explicitly. However, it is easy to derive an analytic formula for the massless case $m_\phi=m_\sigma=0$, which is useful enough to illustrate the similarity of the purity bound and the scattering amplitude bound. In Appendix~\ref{detailcalc}, we derive the following analytic result:
\begin{align}
\label{I_phi_flat}
    I_{\phi}(p) = \left(\frac{\LUV}{f}\right)^4 F^{\text{flat}}_\phi(\pL) \,,
\end{align}
where $F^{\text{flat}}_\phi(\pL)$ is a dimensionless function of $\pL \coloneqq p/\LUV$ and its explicit form is given in~\eqref{C:phi_flat}. Then, the unitarity condition $\gamma_\phi \geq 0$ or equivalently $I_\phi \leq 1$ implies a family of upper bounds on $\LUV/f$:
\begin{align}
\frac{\LUV}{f}  \leq  \left[F^{\text{flat}}_\phi(\pL)\right]^{-\frac{1}{4}}\,.
\end{align}
See Fig.~\ref{fig:UVp_phi_flat} for the bounds as a function of $\pL$. We can optimize the bounds by choosing $\pL$ that maximizes $ F^{\text{flat}}_\phi(\pL)$, which sets the maximum UV cutoff numerically quantified as
\begin{align}
\LUV\leq 20.7\, f\,.
\end{align}
Qualitatively, this reproduces the bound $\LUV \lesssim f$ obtained from the perturbative unitarity of four-point scattering amplitudes. Note that $I_\phi(\pL)$ vanishes, and hence the bound is trivialized in the small $\pL$ limit. This is because the shift symmetry of $\phi$ guarantees that the four-point wavefunction coefficient scales as $\psi_{\phi\phi\sigma\sigma}(\bsp,\bsk_1,\bsk_2,\bsk_3)=\mathcal{O}(p)$ and vanishes in the small $p$ limit.

\subsubsection{Bounds on $\gamma_{\sigma}$}

Next, we consider $\gamma_{\sigma}(p)$, choosing the Fourier modes $\sigma_\bsp$ and $\sigma_{-\bsp}$ as the system. As in the $\gamma_{\phi}(p)$ case, the corresponding purity follows from \eqref{4ptpurity} as
\begin{align}
\gamma_{\sigma}(p)=1-I_{\sigma}(p)
\end{align}
with $I_{\sigma}(p)$ given by
\begin{align}
\label{Is_flat}
    &I_{\sigma}(p)
     = \frac{1}{8} \int_{\bsk_1, \bsk_2}  
   \frac{\left| \psi_{\phi\phi\sigma\sigma}(\bsk_1, \bsk_2, \bsp,\bsk_3) \right|^2}
   {\Re \, \psi_{\phi\phi}(k_1) \Re \, \psi_{\phi\phi}(k_2) \Re \, \psi_{\sigma\sigma}(p) \Re \, \psi_{\sigma\sigma}(k_3)}
 \\
   &=\frac{1}{8 f^4 E^{\sigma}(\bsp)}\int_{\bsk_1, \bsk_2}\frac{1}{
    E^{\phi}(\bsk_1)
    E^{\phi}(\bsk_2)
    E^{\sigma}(\bsk_3)} \left|\frac{E^{\phi}(\bsk_1)E^{\phi}(\bsk_2) -\bsk_1\cdot\bsk_2}{E^{\phi}(\bsk_1) + E^{\phi}(\bsk_2) + E^{\sigma}(\bsk_3) + E^{\sigma}(\bsp)}\right|^2\,.
    \notag
\end{align}
As before, $\bsk_3=-(\bsp+\bsk_1+\bsk_2)$ and the integral region of $\bsk_{1,2}$ is defined such that
\begin{align}
0\leq E^\phi(\bsk_1),E^\phi(\bsk_2),E^\sigma(\bsk_3),E^\sigma(\bsp)\leq\LUV
\,,
\end{align}
where we set $\LIR=0$ since the integral is IR-finite.

\begin{figure}[t]
    \centering
    \includegraphics[height=7cm]{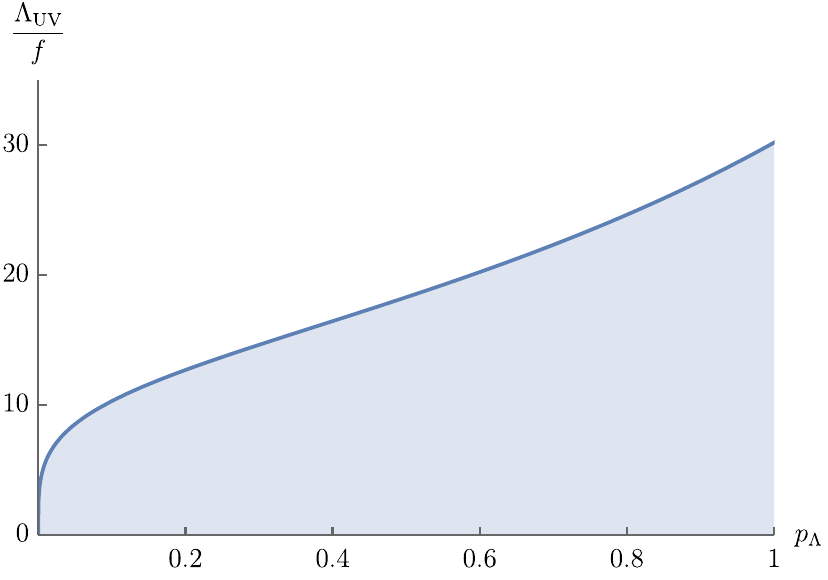}
    \caption{
     The allowed region for $\LUV/ f$ as a function of $\pL$, shown as the shaded area, is derived from the unitarity bound imposed by the purity $\gamma_{\sigma}$ in flat spacetime.
    }
    \label{fig:UVp_sigma_flat}
\end{figure}

Similarly to the $I_\phi(p)$ case, we can perform the integral analytically if $m_\phi=m_\sigma=0$. The result is summarized schematically as
\begin{align}
    I_{\sigma}(p)= \left(\frac{\LUV}{f}\right)^4 F^{\text{flat}}_\sigma(\pL)\,,
\end{align}
where $F^{\text{flat}}_\sigma(\pL)$ is a dimensionless function of $\pL=p/\LUV$. See \eqref{C:sig_flat} for its explicit form. Then, the unitarity condition $\gamma_\sigma \geq 0$ or equivalently $I_\sigma \leq 1$ implies a family of upper bounds on $\LUV/f$:
\begin{align}
\frac{\LUV}{f} \leq \left[F^{\text{flat}}_\sigma(\pL) \right]^{-\frac{1}{4}}\,.
\end{align}
See Fig.~\ref{fig:UVp_sigma_flat} for the bounds as a function of $\pL$, which shows that the upper bound increases monotonically with increasing $\pL$. In particular, it vanishes at $\pL=0$. Since the bounds have to be satisfied for all $p$, this implies that $\LUV=0$ and therefore there is no regime of validity of the EFT.

Actually, this breakdown of EFT is just an artifact of our parameter choice $m_\sigma=0$. Instead, a lower bound on the mass $m_\sigma$ of $\sigma$ can be derived by studying the case $m_\sigma\neq0$.
For this, let us first remind that $E^\sigma(\bsp)=p$ in the massless limit $m_\sigma=0$ and hence the prefactor in the second line of~\eqref{Is_flat} gives a divergence $1/E^\sigma(p)=1/p$ in the soft limit $p\to0$.  
Note that the remaining momentum integral is finite even in this limit.
Around this limit, $F_\sigma^{\text{flat}}(\pL)$ is expanded as
\begin{align}
F_\sigma^{\text{flat}}(\pL)=\frac{1}{64\,(2\pi)^4\,\pL} \left[1 + \mathcal{O}\left(\pL\right)\right]\,,
\end{align}
and accordingly, for a generic mass $m_\sigma\neq0$ well below the UV cutoff $m_\sigma\ll\LUV$, we find 
\begin{align}
F_{\sigma,\,{\rm massive}}^{\text{flat}}(0)=\frac{1}{64\,(2\pi)^4}\frac{\LUV}{m_\sigma}\left[1+\mathcal{O}\left(\frac{m_\sigma}{\LUV}\right)\right]
\,.
\end{align}
Therefore, the bound $I_\sigma(0)\leq1$ implies a lower bound on $m_\sigma$:
\begin{align}
\label{lower_bound_mass}
\frac{m_\sigma}{\LUV}
\geq \frac{1}{64\,(2\pi)^4}\left(\frac{\LUV}{f}\right)^4\,.
\end{align}
See also Fig.~\ref{fig:Fp_flat} in Appendix~\ref{ExpPurF} for a concrete profile of $F_\sigma^{\text{flat}}(p)$ in the massive $\sigma$ case. Note that $I_\sigma(p)\leq1$ with $\pL=\mathcal{O}(0.1)$ gives
$\LUV\lesssim f$,
similarly to the $\gamma_\phi$ case.

\section{Perturbative unitarity bounds in de Sitter spacetime}\label{sec:deSitter}

In this section, we extend the analysis to de Sitter spacetime:
\begin{align}
    \dd s^2 = \frac{- \dd \eta^2 + \dd \bsx^2}{H^2 \eta^2}\,,
\end{align}
which corresponds to the scale factor $a(\eta)=-1/(H\eta)$ in \eqref{hom_iso}. As before, we perform detailed analysis in the two-scalar model,
\begin{align}
\label{two-scalar-model-dS}
S&=\int \dd^4x\sqrt{-g}\left[
-\frac{1}{2}(\partial_\mu \phi)^2
-\frac{1}{2}m_\phi^2\phi^2
-\frac{1}{2}(\partial_\mu\sigma)^2
-\frac{1}{2}m_\sigma^2\sigma^2
+\frac{\epsilon}{4f^2} \sigma^2(\partial_\mu\phi)^2
\right]
\,,
\end{align}
and discuss the impacts of the Hubble scale $H$ on the perturbative unitarity bounds. A qualitative discussion on general multi-scalar models is given at the end of the section.

\subsection{Illustrative example: \texorpdfstring{$m_\phi=0$}{massless phi} and \texorpdfstring{$m_\sigma=\sqrt{2}H$}{conformal mass sigma}}

As an illustrative example, we first consider the two-scalar model~\eqref{two-scalar-model-dS} and assume that $\phi$ is massless and $\sigma$ has a conformal mass:
\begin{align}
    m_\phi = 0\,, \quad m_\sigma=\sqrt{2}H\,.
\end{align}
For this parameter choice,  the bulk-to-boundary propagators have a simple form (see also Appendix~\ref{BtoBp}),
\begin{align}
    K_\phi (k;\eta)=\frac{1-ik\eta}{1-ik\eta_*}\,e^{ik(\eta-\eta_*)}\,,\quad K_\sigma (k;\eta)=\frac{\eta}{\eta_*}\,e^{ik(\eta-\eta_*)}\,,
\end{align}
and all the detailed analysis can be performed analytically.

If we define the wavefunction coefficients as~\eqref{wavefunc_two-scalar} in the same manner as the flat spacetime case, the two-point coefficients follow from the general formula~\eqref{2pt_wave} as
\begin{align}
    \Re \, \psi_{\phi \phi}(k;\eta_*) = \frac{k^3}{H^2 \left(1 + k^2 \eta_*^2\right)}\,,\quad \Re \, \psi_{\sigma \sigma}(k;\eta_*)= \frac{k}{H^2 \eta_*^2}\,.
\label{psi2_dS}
\end{align} 
Similarly, using~\eqref{psi4_general}, the four-point coefficient reads
\begin{align}
\notag
&\psi_{\phi\phi\sigma\sigma}(\bsk_1,\bsk_2,\bsk_3,\bsk_4;\eta_*)\\
\notag
& = -\frac{\epsilon k_1 k_2}{f^2 H^2 \eta_*^2 (1-i k_1 \eta_*) (1-i k_2 \eta_*) k_T^3}\\
& \quad \times \bigg[k_1 k_2 \Big(1+(1-i k_T \eta_*)^2\Big) (1-\cos\theta) - \Big((k_1+k_2) \left(1-i k_T \eta_* \right) + k_T\Big) k_T \cos\theta\bigg]\,,
\label{psi4_dS}
\end{align}
where $k_T \coloneqq k_1 + k_2 + k_3 + k_4$ and the angle $\theta$ is defined such that $\bsk_1 \cdot \bsk_2 = k_1 k_2 \cos\theta\,$.

Now we are ready to compute the purity in the same manner as~\eqref{purityphiflat} and~\eqref{Is_flat}. Similarly to the flat space case, the momentum integral turns out to be IR-finite, so that we set $\LIR=0$. On the other hand, the UV cutoff $\LUV$ gives an upper bound on the physical momentum $k_{\text{phys}}$ (rather than the comoving momentum $k$) as
\begin{align}
    k_\text{phys} :=\frac{k}{a(\eta_*)} =-kH\eta_* \leq \LUV\,.
\end{align}
Note that the UV cutoff scale has to be well above the Hubble scale $\LUV\gtrsim H$, otherwise the horizon-scale dynamics cannot be captured.
Below, we present the bounds on $\gamma_\phi$ and $\gamma_\sigma$, and their implications in order.

\subsubsection{Bounds on $\gamma_\phi$}

First, we choose the Fourier modes $\phi_\bsp$ and $\phi_{-\bsp}$ as the system. Then, the purity $\gamma_{\phi}(p)$ can be computed analytically by substituting~\eqref{psi2_dS}-\eqref{psi4_dS} into the first line of~\eqref{purityphiflat} and performing the momentum integral in the same manner as in Appendix~\ref{detailcalc}. The result is schematically given as an expansion in $H$ as
\begin{align}
\gamma_\phi(p) = 1 -I_\phi(p)
\end{align}
with $I_\phi(p)$ given by
\begin{align}
\label{I_phi_dS}
I_\phi(p)= \left(\frac{\LUV}{f}\right)^4 F_{\phi0}^{\text{dS}}\left(\bpL \right) + \left(\frac{H}{f}\right)^2 \left(\frac{\LUV}{f}\right)^2 F_{\phi2}^{\text{dS}} \left(\bpL \right) +\left(\frac{H}{f}\right)^4 F_{\phi4}^{\text{dS}}\left(\bpL \right)\,,
\end{align}
where the coefficients, $F_{\phi0}^{\text{dS}}(\bpL),\, F_{\phi2}^{\text{dS}}(\bpL)$, and $F_{\phi4}^{\text{dS}}(\bpL)$, of each order of $H$ are dimensionless functions of the ratio $\bpL \coloneqq p_{\text{phys}}/\LUV$ of the physical momentum $p_{\text{phys}}\coloneqq -pH\eta_*$ and the UV cutoff $\LUV$. Their explicit forms are given in \eqref{C:phi0_dS}-\eqref{C:phi4_dS}. We emphasize that $\gamma_\phi$ does not depend on $\eta_*$ explicitly once we write it in terms of the physical momentum $p_{\mathrm{phys}}$, as a consequence of the de Sitter dilatation symmetry. Another point to notice is that the $\mathcal{O}(H^0)$ term reproduces the flat spacetime result:
\begin{align}
     F_{\phi0}^{\text{dS}}(\bpL) = F_\phi^{\text{flat}}(\pL) \Big|_{\pL = \bpL}\,.
\end{align}
In particular, as expected, the flat spacetime approximation works well as long as the UV cutoff $\LUV$ is well above the Hubble scale (for fixed $\bpL$).

\begin{figure}[t]
    \centering
    \includegraphics[height=7cm]{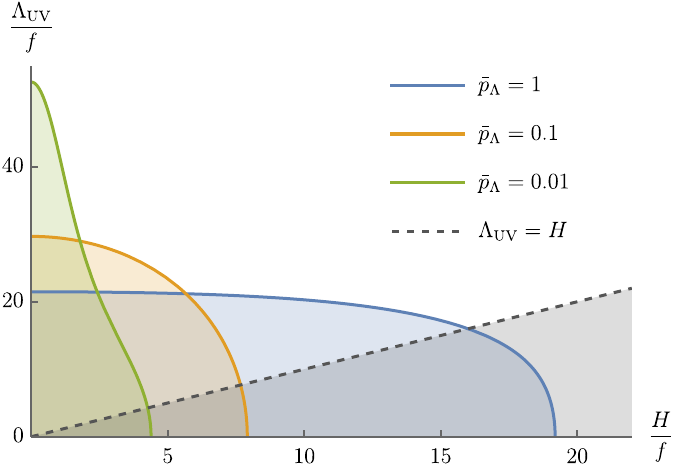}
    \caption{
    The allowed region for $\LUV/ f$ and $H/f$ as a function of $\bpL$, shown as the shaded area, is derived from the unitarity bound imposed by the purity $\gamma_{\phi}(\bpL)$ in de Sitter spacetime. The gray region indicates $\LUV \le H$, which is beyond our interests.
    }
    \label{fig:UVH_phi_dS}
\end{figure}

Then, the unitarity condition $\gamma_\phi \geq 0$ implies $I_\phi \leq 1$ with~\eqref{I_phi_dS}.
For given $\bpL$, this gives a bound on the two parameters, $\LUV/f$ and $H/f$. Fig.~\ref{fig:UVH_phi_dS} shows the allowed regions for $\bpL=1,0.1,0.01$. For a representative choice $\bpL=\mathcal{O}(0.1)$, the bounds imply the following two conditions:
\begin{align}
\LUV\lesssim f\,,
\quad
H\lesssim f\,.
\end{align}
The first condition is analogous to the flat space result, which shows that the maximum UV cutoff is around the field-space curvature scale. The second condition is a consequence of the thermal nature of de Sitter spacetime: de Sitter spacetime has a temperature $T=H/(2\pi)$ and the temperature cannot exceed the maximum UV cutoff $\sim f$ of the theory. More quantitatively, we find that the maximum UV cutoff for fixed $\bpL$ monotonically decreases with increasing $H$, vanishing at a maximum Hubble scale.

In addition, the allowed region disappears in the limit $\bpL\to0$, which corresponds to the superhorizon limit of the system modes $\phi_\bsp$ and $\phi_{-\bsp}$. To elaborate on this point, it is convenient to note the small $\bpL$ behavior of $I_\phi(p)$:
\begin{align}
I_\phi(p) = \frac{1}{(2\pi)^4 \, \bpL} \left[\frac{23-54 \log \left(\frac{3}{2}\right)}{216} \left(\frac{H}{f}\right)^2 \left(\frac{\LUV}{f}\right)^2 + \frac{101-204 \log \left(\frac{3}{2}\right)}{432} \left(\frac{H}{f} \right)^4 + \mathcal{O}(\bpL)\right]\,.
\end{align}
\begin{figure}[t]
    \centering
    \includegraphics[height=7cm]{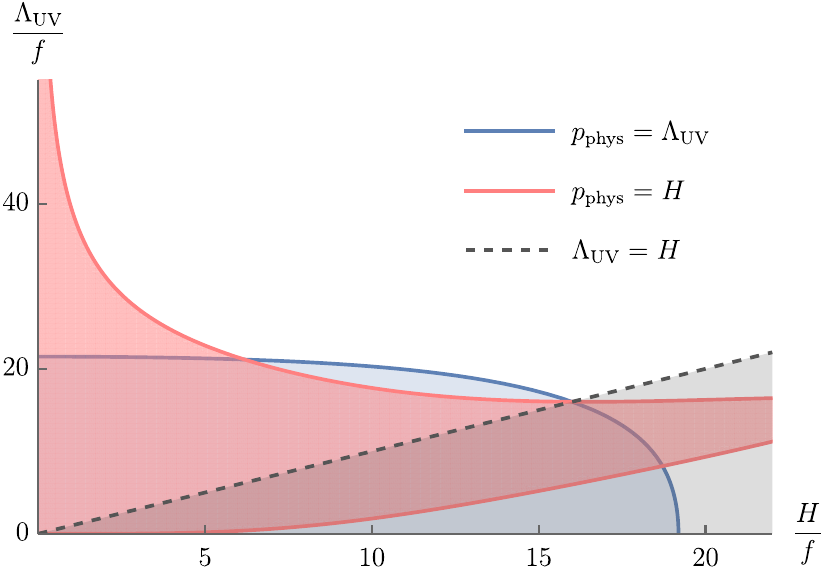}
    \caption{
    The allowed region for $\LUV/ f$ and $H/f$: those for $p_\mathrm{phys} = H$ and $p_\mathrm{phys} = \LUV$ are illustrated. Their overlap satisfies the bounds for all $p_{\mathrm{phys}}$ within the range $H \le p_\mathrm{phys} \le \LUV$.
    }
    \label{fig:UVH_phi_dS_Hubble}
\end{figure}
First, we notice that the singular behavior appears only when $H\neq0$. In fact, as we discussed in Sec.~\ref{subsec:purity_UV_flat}, $I_\phi(p)$ on flat spacetime vanishes and thus the bound is trivialized in the limit $p\to0$. See also Fig.~\ref{fig:UVp_phi_flat}. A similar singularity in the limit $p\to0$ on de Sitter spacetime was pointed out in~\cite{Pueyo:2024twm} and perturbative unitarity bounds on squeezed configurations were studied. In the next subsection, we argue that such singular behaviors are peculiar to light fields in the complementary series, which are tachyonic at the superhorizon scale.

It should be emphasized that the singularity in the IR limit $p\to0$ mentioned above does not necessarily imply breakdown of the EFT, but rather indicates breakdown of the perturbative computation of purity. Indeed, late-time correlators in this setup are IR finite, and perturbative computation works well without suffering from tachyonic behavior at the superhorizon scale. This suggests that for light fields, purity does not offer an appropriate ground to study perturbative unitarity bounds for cosmological perturbations.
\footnote{Notice, however, that the unitarity bound on purity has to be respected at the non-perturbative level, even if its perturbative computation breaks down.}
It would be important to explore other observables more appropriate for this purpose, leaving it for future work.
Nevertheless, with this caveat in mind, we expect that purity can still be used to estimate the UV cutoff scale of the EFT by choosing the system mode inside the horizon $p_\mathrm{phys} \gtrsim H$, for which purity is free from the IR divergence peculiar to purity of light fields. In Fig.~\ref{fig:UVH_phi_dS_Hubble}, we illustrate the bounds evaluated at $p_\mathrm{phys} = H$ and $p_\mathrm{phys} = \LUV$ in order to identify the region of parameter space that remains allowed for all momenta within the range $H \le p_\mathrm{phys} \le \LUV$. There, we find that $f$ has to be bigger than $\LUV$ and $H$, similarly to the massive case. The same remark applies to $\gamma_\sigma$ as well.

\subsubsection{Bounds on $\gamma_\sigma$}

Next, we choose the Fourier modes $\sigma_\bsp$ and $\sigma_{-\bsp}$ as the system. The purity $\gamma_{\sigma}(p)$ can be computed analytically in the same manner as before:
\begin{align}
\gamma_\sigma(p) = 1-I_\sigma(p)
\end{align}
with $I_\sigma(p)$ given by
\begin{align}
I_\sigma(p)= \left(\frac{\LUV}{f}\right)^4 F_{\sigma0}^{\text{dS}}\left(\bpL \right) + \left(\frac{H}{f}\right)^2 \left(\frac{\LUV}{f}\right)^2 F_{\sigma2}^{\text{dS}} \left(\bpL \right) +\left(\frac{H}{f}\right)^4 F_{\sigma4}^{\text{dS}}\left(\bpL \right)\,.
\end{align}
See \eqref{C:sig0_dS}-\eqref{C:sig4_dS} for an explicit form of the coefficients, $F_{\sigma0}^{\text{dS}}\left(\bpL \right)$, $F_{\sigma2}^{\text{dS}}\left(\bpL \right)$ and $F_{\sigma4}^{\text{dS}}\left(\bpL \right)$, of the expansion in $H$. In particular, the $\mathcal{O}(H^0)$ coefficient reproduces the flat space result:
\begin{align}
     F^{\text{dS}}_{\sigma0}(\bpL)
    & = F^{\text{flat}}_\sigma(\pL) \Big|_{\pL = \bpL}\,.
\end{align}
\begin{figure}[t]
    \centering
    \includegraphics[height=7cm]{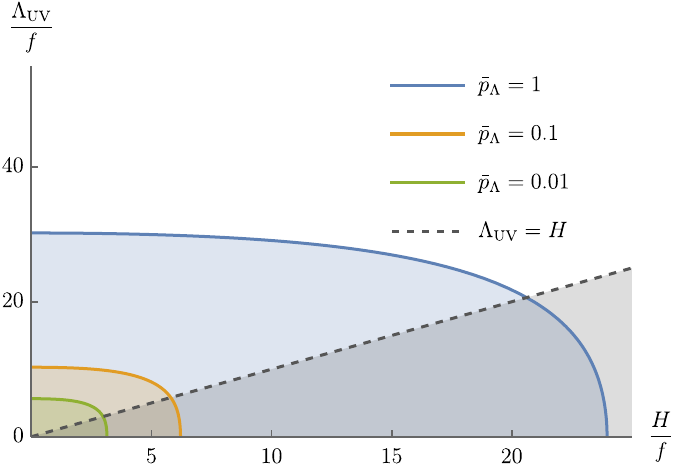}
    \caption{
    The allowed region for $\LUV/ f$ and $H/f$ as a function of $\bpL$, shown as the shaded area, is derived from the unitarity bound imposed by the purity $\gamma_{\sigma}(\bpL)$ in de Sitter spacetime. The gray region indicates $\LUV \le H$, which is beyond our interests.
    }
    \label{fig:UVH_sig_dS}
\end{figure}
In Fig.~\ref{fig:UVH_sig_dS}, the regions compatible with the bound $\gamma_\sigma(p)\geq0$ ($I_\sigma(p)\leq1$) are shown for $\bpL=1,0.1,0.01$. Similarly to the $\gamma_\phi$ case, the bounds for $\bpL=\mathcal{O}(0.1)$ imply
\begin{align}
f\gtrsim H\,,
\quad
\LUV\lesssim f\,.
\end{align}
Also, the allowed region shrinks in the soft limit $\bpL\to0$, which can be confirmed explicitly from the small $\bpL$ behavior of $I_\sigma(p)$:
\begin{align}
I_\sigma(p) = \frac{1}{(2\pi)^4 \, \bpL} \Bigg[\frac{1}{64} \left(\frac{\LUV}{f}\right)^4 + & \frac{-9+24\log \left(\frac{3}{2}\right)}{32} \left(\frac{H}{f}\right)^2 \left(\frac{\LUV}{f}\right)^2 \notag\\
& \hspace{40pt} + \frac{61-120 \log \left(\frac{3}{2}\right)}{72}\left(\frac{H}{f} \right)^4 + \mathcal{O}(\bpL)\Bigg]\,.
\end{align}
As we discuss in the next subsection, this singular behavior is due to our parameter choice $m_\sigma=\sqrt{2}H$, for which $\sigma$ is in the complementary series. Note that the singularity of $I_\sigma$(p) persists even in the flat-space limit, which is consistent with the analysis in Sec.~\ref{subsec:purity_UV_flat}.

\subsection{More on superhorizon limit}

In the previous subsection, we encountered divergence of the purity $\gamma_{\phi,\sigma}(p)$ in the superhorizon limit $p\to0$ of the system modes.
In this subsection, we elaborate on this point and argue that this divergence is peculiar to light fields in the complementary series.

\paragraph{Light and heavy fields in de Sitter spacetime.}

To identify the origin of divergence in the purity, let us first recall the superhorizon behavior of scalar fields in de Sitter spacetime. In terms of the physical time coordinate $t=-H^{-1}\ln (-H\eta)$, the superhorizon behavior of the equation of motion for a free scalar $\varphi$ of mass $m$ reads
\begin{align}
\label{eom_varphi}
\ddot{\varphi}_\bsk+3H\dot{\varphi}_\bsk+m^2\varphi_\bsk\simeq 0
\quad
\left(\frac{k_{\text{phys}}}{H}=|k\eta|\ll1\right)\,,
\end{align}
where the dots stand for the derivatives in $t$. When the mass is in the range $0\leq m<\frac{3}{2}H$, i.e., when the scalar $\varphi$ is in the complementary series, the solution to the equation of motion~\eqref{eom_varphi} is overdamped:
\begin{align}
\varphi_\bsk\sim e^{-\frac{3}{2} Ht}e^{\pm \nu Ht}
\quad
\text{with}
\quad
\nu=\sqrt{\frac{9}{4}-\frac{m^2}{H^2}}\,.
\end{align}
On the other hand, when $m>\frac{3}{2}H$, i.e., when $\varphi$ is in the principal series, \eqref{eom_varphi} accommodates oscillating solutions:
\begin{align}
\varphi_\bsk\sim e^{-\frac{3}{2}Ht}e^{\pm i\mu Ht}
\quad
\text{with}
\quad
\mu=\sqrt{\frac{m^2}{H^2}-\frac{9}{4}}\,.
\end{align}
Hence, from the superhorizon perspective, the light scalars ($0\leq m<\frac{3}{2}H$) in the complementary series can be regarded as tachyonic fields.

\paragraph{Two-point wavefunction coefficient.}

The mass dependence of the superhorizon behavior explained above is well captured by the two-point wavefunction coefficient $\psi_{\varphi\varphi}$. First, for the heavy scalars ($m>\frac{3}{2}H$) in the principal series, the real part of the two-point wavefunction $\psi_{\varphi\varphi}$ in the superhorizon limit $|k\eta_*|\ll1$ follows from the general formula~\eqref{2pt_wave} and the bulk-to-boundary propagator~\eqref{K_heavy} of the Bunch-Davies vacuum as
\begin{align}
\Re\,\psi_{\varphi\varphi}(k,\eta_*)&\simeq
\frac{2}{1+\coth \pi\mu}\,a(\eta_*)^3\,
\mu H
\nonumber
\\
\label{psi_phiphi_heavy}
&\quad
\times
\left[
1
+e^{-\pi\mu}
e^{i\alpha(\mu)}(-k\eta_*)^{2i\mu}
\right]^{-1}
\left[
1
+e^{-\pi\mu}
e^{-i\alpha(\mu)}(-k\eta_*)^{-2i\mu}
\right]^{-1}\,,
\end{align}
where $\alpha(\mu)$ is a mass-dependent phase factor defined in~\eqref{K_heavy}.
Note that the factor $e^{-\pi\mu}$ in front of the oscillating terms $(-k\eta_*)^{\pm 2i\mu}$ in the second line corresponds to the square root of the Boltzmann factor $e^{-2\pi\mu}$, which reflects the thermal nature of the Bunch-Davies vacuum.
Indeed, in the heavy-mass limit $\mu\gg1$, thermal particle creation is exponentially suppressed, and we have
\begin{align}
\Re\,\psi_{\varphi\varphi}(k,\eta_*)&\simeq
\,a(\eta_*)^3\,
\mu H
\qquad
\text{for}
\qquad
|k\eta|\ll1,\quad\mu\gg1
\,,
\end{align}
which coincides with the flat space wavefunction~\eqref{2pt_flat} up to an overall normalization volume factor $a(\eta_*)^3$ under the identification of the energy $E=\mu H$. For generic $\mu$ (except $\mu\ll1$), the superhorizon behavior reads
\begin{align}
\Re\,\psi_{\varphi\varphi}(k,\eta_*)&=\mathcal{O}(1)\times
a(\eta_*)^3\mu H
\qquad
\text{for}
\qquad
|k\eta|\ll1,\quad\mu\gtrsim 1
\,,
\end{align}
which is finite in particular.

In contrast, for the light scalars $(0\leq m<\frac{3}{2}H)$, the two-point wavefunction coefficient $\psi_{\varphi\varphi}$ vanishes in the superhorizon limit $|k\eta|\ll1$:
\begin{align}
\Re\,\psi_{\varphi\varphi}(k,\eta_*)\simeq \frac{2^{1-2\nu}\,\pi}{\Gamma(\nu)^2}\,a(\eta_*)^3\,H\,(-k\eta_*)^{2\nu}
\,,
\end{align}
which signals tachyonic enhancement of superhorizon fluctuations.

\paragraph{Purity.}

Finally, we discuss the superhorizon limit of the four-point wavefunction coefficients~\eqref{psi4_general} and the purity. First, in the superhorizon limit $p\to0$ of $\phi$, which has a derivative coupling, the four-point wavefunction coefficient $\psi_{\phi\phi\sigma\sigma}$ scales as
\begin{align}
\psi_{\phi\phi\sigma\sigma}(\bsp,\bsk_2,\bsk_3,\bsk_4)
=\left\{\begin{array}{cc}\mathcal{O}(p^0) & (m_\phi\neq0)\,, \\[1mm]
\mathcal{O}(p^1)&(m_\phi=0)\,. 
\end{array}\right.
\end{align}
On the other hand, in the superhorizon limit $p\to0$ of $\sigma$, $\psi_{\phi\phi\sigma\sigma}$ scales as
\begin{align}
\psi_{\phi\phi\sigma\sigma}(\bsk_1,\bsk_2,\bsp,\bsk_4)
=\mathcal{O}(p^0)\,.
\end{align}
Then, in the superhorizon limit $p\to0$, the purity scales as
\begin{align}
\gamma_{\phi}(p)=\left\{\begin{array}{ll}
\mathcal{O}(p^{0}) &\,\, (m_\phi>\frac{3}{2}H)\,, \\[1mm]
\mathcal{O}(p^{-2\nu_\phi}) & \,\,(0<m_\phi<\frac{3}{2}H)\,, \\[1mm]
\mathcal{O}(p^{-1})&\,\,(m_\phi=0)\,. 
\end{array}\right.
\quad
\gamma_{\sigma}(p)=\left\{\begin{array}{ll}
\mathcal{O}(p^{0}) &\,\, (m_\sigma>\frac{3}{2}H)\,, \\[3mm]
\mathcal{O}(p^{-2\nu_\sigma}) & \,\,(0\leq m_\sigma<\frac{3}{2}H)\,. 
\end{array}\right.
\end{align}
We conclude that all the divergences we encountered in the superhorizon limit $p\to0$ are due to vanishing two-point coefficients $\psi_{\phi\phi},\psi_{\sigma\sigma}$ that reflect the tachyonic superhorizon behavior of light fields in the complementary series. In particular, such divergences are absent if we consider massive fields in the primary series.

\subsection{Extension to \texorpdfstring{$N$}{N}-scalar model}

Finally, we discuss the extension to general $N$-scalar models qualitatively. For this, it is convenient to work in the Riemann normal coordinates of the field space:
\begin{align}
G_{IJ}(\phi)=\delta_{IJ}-\sum_{K,L}\frac{1}{3}R_{IKJL}\,\phi^K\phi^L+\mathcal{O}(\phi^3)\,,
\end{align}
where $R_{IKJL}$ is the Riemann tensor of the field space evaluated at the origin $\phi=0$.
The corresponding four-point coupling $C_{IJKL}$ defined in~\eqref{nonlinear_S} reads
\begin{align}
C_{IJKL}=\frac{1}{3}R_{IKJL}\,.
\end{align}
If we choose the modes $\phi^{\bar{I}}_{\pm\bsp}$ of the field index $\bar{I}$ and the comoving momentum $\pm\bsp$ as the system, the purity can be computed using the general formula~\eqref{psi4_general}. 

For illustration, let us first consider the case where all the scalar fields have the same mass $m$. In this simple setup, the wavefunction coefficients are schematically of the form,
\begin{align}
\psi_{IJ}(k)=\delta_{IJ}\,\bar{\psi}_2(k)\,,
\quad
\psi_{IJKL}(\bsk_1,\bsk_2,\bsk_3,\bsk_4)=R_{IKJL}\bar{\psi}_4(\bsk_1,\bsk_2,\bsk_3,\bsk_4)\,,
\end{align}
with momentum-dependent factors $\bar{\psi}_2(k)$ and $\bar{\psi}_4(\bsk_1,\bsk_2,\bsk_3,\bsk_4)$ that do not carry the species index.
Then, the purity reads
\begin{align}
\gamma_{\phi^{\bar{I}}}(p)=1-\mathcal{I}(p)\sum_{J,K,L}R_{\bar{I}KJL}^2
\end{align}
with $\mathcal{I}(p)$ given by
\begin{align}
    \mathcal{I}(p)
    &= \frac{1}{24}\int_{\bsk_1, \bsk_2}    \frac{\left|\bar{\psi}_4(\bsp, \bsk_1, \bsk_2,\bsk_3) \right|^2}
   {\Re \,\bar{\psi}_2(p) \,\Re\, \bar{\psi}_2(k_1) \,\Re \,\bar{\psi}_2(k_2) \,\Re \,\bar{\psi}_2(k_3)}\,,
\end{align}
where $\bsk_3=\bsp-\bsk_1-\bsk_2$ and the integral region is $0\leq -k_i\eta_*H \leq \LUV$ ($i=1,2,3$) as before. Now the perturbative unitarity bound implies the following upper bound of the Riemann tensor squared with three indices contracted:
\begin{align}
\sum_{J,K,L}R_{\bar{I}JKL}^2\leq\frac{1}{\mathcal{I}(p)}\,.
\end{align}
In particular, for $m\sim H$ and $\bpL=p_{\rm phys}/\LUV=\mathcal{O}(0.1)$, we have $\mathcal{I}(p)\sim \LUV^4$, so that the bound~is
\begin{align}
\sum_{J,K,L}R_{\bar{I}JKL}^2
\lesssim\frac{1}{\LUV^4}\,.
\end{align}
Note that the above result holds more generally even beyond the identical mass case, as long as $p_{\rm phys},\LUV\gg m_I,H$. On the other hand, if the system modes are at the superhorizon scale, the bounds are sensitive to the masses. For example, if the system mode is a scalar in the complementary series, the purity diverges at the superhorizon scale as a consequence of tachyonic behavior.

\section{Conclusion}
\label{sec:conclusions}

In this paper, we studied perturbative unitarity bounds on the field-space curvature in de Sitter spacetime, using the momentum-space entanglement approach recently proposed in~\cite{Pueyo:2024twm}. We first analyzed purity in flat spacetime and showed that the UV cutoff is set by the field-space curvature, in agreement with results from the amplitude analysis. We then extended the analysis to de Sitter spacetime, where we derived unitarity bounds that involve not only the UV cutoff and the field-space curvature, but also the Hubble scale. Unlike the original paper~\cite{Pueyo:2024twm}, our analysis was performed without taking the superhorizon limit, which allowed us to interpolate the flat space analysis and the superhorizon analysis. In particular, we derived an upper bound on the field-space curvature of the Hubble scale order, which reflects the thermal nature of de Sitter spacetime, in addition to a bound analogous to the flat space result. We also provided a detailed discussion on the superhorizon behavior of purity, interpreted in terms of the tachyonic/non-tachyonic superhorizon behavior of light/heavy fields in the complementary/principal series.

To conclude, we outline several interesting directions for future work. First, it would be worthwhile to extend our analysis to more realistic inflationary models such as Higgs inflation~\cite{Bezrukov:2007ep} and quasi-single field inflation~\cite{Chen:2009zp}, for which perturbative unitarity bounds were previously studied under the flat space approximation~\cite{Lerner:2010mq,Giudice:2010ka,Atkins:2010yg,Calmet:2013hia,Barbon:2015fla,Ema:2020zvg,Kim:2021pbr}. Second, it is important to go beyond the perturbative analysis of unitarity bounds. For this, it would be useful to investigate purity in models with phase transitions, where the UV completion is achieved in a non-perturbative manner. Finally, it would be crucial to study analyticity of purity and to perform partial wave type expansions, with the aim of formulating a bootstrap program based on entanglement measures. We hope to revisit these issues in the near future.

\subsection*{Acknowledgement}

K.N. is supported in part by JSPS KAKENHI Grant Number JP22J20380. T.N. is supported in part by JSPS KAKENHI Grant No. JP22H01220 and MEXT KAKENHI Grant No. JP21H05184 and No. JP23H04007.

\newpage
\appendix

\section{Bulk-to-boundary propagator in de Sitter spacetime}\label{BtoBp}

This appendix summarizes properties of the scalar bulk-to-boundary propagators in de Sitter spacetime for general masses. In de Sitter spacetime, the equation of motion for a free scalar $\varphi_{\bsk}(\eta)$ in the (spatial) Fourier space is given by
\begin{align}
\left[\theta_\eta(\theta_\eta-3)+\frac{m^2}{H^2}+k^2\eta^2\right]\varphi_{\bsk}(\eta)=0\,,
\end{align}
where the Euler operator $\theta_\eta=\eta\partial_\eta$ counts the exponent of the conformal time $\eta$, $H$ is the constant Hubble parameter, $m$ is the mass of the scalar field $\varphi$, $\bsk$ is the comoving spatial momentum, and $k=|\bsk|$. The bulk-to-boundary propagator $K_\varphi (k;\eta)$ of the Dirichlet problem is defined as a solution to the free equation of motion,
\begin{align}
\left[\theta_\eta(\theta_\eta-3)+\frac{m^2}{H^2}+k^2\eta^2\right]K_\varphi (k;\eta)=0\,,
\end{align}
that satisfies the boundary conditions,
\begin{align}
K_\varphi (k;\eta_*)=1\,,
\quad
\lim_{\eta\to-(1-i\epsilon)\infty}K_\varphi (k;\eta)=0\,,
\end{align}
where $\eta_*$ is the conformal time at which the wavefunction is evaluated. The second condition is the Bunch-Davies vacuum condition. As we see shortly, its properties are qualitatively different between light scalars ($0\leq m<\frac{3}{2}H$) in the complementary series and heavy scalars ($m>\frac{3}{2}H$) in the principal series, so that we discuss the two cases separately in the following.

\paragraph{Light fields.}

For light scalars $0\leq m<\frac{3}{2}H$, the bulk-to-boundary propagator reads
\begin{align}
K_\varphi (k;\eta)=\frac{(-\eta)^{\frac{3}{2}}H_{\nu}^{(2)}(-k\eta)}{(-\eta_*)^{\frac{3}{2}}H_{\nu}^{(2)}(-k\eta_*)}
\quad
\text{with}
\quad
\nu=\sqrt{\frac{9}{4}-\frac{m^2}{H^2}}
\,,
\end{align}
where $H_\nu^{(2)}(z)$ is the Hankel function of the second kind. Note that it simplifies for the massless case as
\begin{align}
K_\varphi (k;\eta)=\frac{1-ik\eta}{1-ik\eta_*}\,e^{ik(\eta-\eta_*)}
\quad (m=0)\,,
\end{align}
and also for the conformal mass case as
\begin{align}
K_\varphi (k;\eta)= \frac{\eta}{\eta_*}\,e^{ik(\eta-\eta_*)}\quad \left(m=\sqrt{2}H\right)\,.
\end{align}
In the superhorizon limit $-k\eta\ll 1$, the bulk-to-boundary propagator enjoys a power-law,
\begin{align}
K_\varphi (k;\eta)\simeq \left(\frac{\eta}{\eta_*}\right)^{\frac{3}{2}-\nu}
\,.
\end{align}

\paragraph{Heavy fields.}

For heavy scalars $m >\frac{3}{2}H$, the bulk-to-boundary propagator is given by
\begin{align}
K_\varphi (k;\eta)=\frac{(-\eta)^{\frac{3}{2}}H_{-i\mu}^{(2)}(-k\eta)}{(-\eta_*)^{\frac{3}{2}}H_{-i\mu}^{(2)}(-k\eta_*)}
\quad
\text{with}
\quad
\mu=\sqrt{\frac{m^2}{H^2}-\frac{9}{4}}
\,.
\end{align}
In contrast to the light scalar case, the subscript of the Hankel function is pure imaginary. As a consequence, its superhorizon behavior is not a simple power-law:
\begin{align}
\label{K_heavy}
K_\varphi (k;\eta)\simeq \left(\frac{\eta}{\eta_*}\right)^{\frac{3}{2}-i\mu}
\frac{1+e^{-\pi\mu}e^{i\alpha(\mu)}(-k\eta)^{2i\mu}}{1+e^{-\pi\mu}e^{i\alpha(\mu)}(-k\eta_*)^{2i\mu}}
\quad
\text{with}
\quad
e^{i\alpha(\mu)}=\frac{\Gamma(-i\mu)}{2^{2i\mu}\Gamma(i\mu)}\,.
\end{align}
Here, we introduced a mass-dependent phase factor $\alpha(\mu)$. The first and second terms in the numerator describe the positive and negative frequency modes at late time, respectively. The prefactor $e^{-\pi\mu}e^{i\alpha(\mu)}$ of the negative frequency mode is nothing but the one appearing in the thermal Bogoliubov coefficients, which reflects the thermal nature of de Sitter spacetime. In particular, in the heavy-mass limit $m\gg H$, only the positive frequency mode remains, and the propagator is reduced to the vacuum one:
\begin{align}
K_\varphi (k;\eta)\simeq \left(\frac{\eta}{\eta_*}\right)^{\frac{3}{2}-i\mu}
\quad
(-k\eta\ll 1\,,
\,\,
m\gg H)\,.
\end{align}

\section{Details of purity computation}\label{detailcalc}

This appendix provides details of the purity computation, especially on the momentum integrals in~\eqref{purityphiflat}. For illustration, we present the analysis of $\gamma_\phi$ on flat spacetime, but $\gamma_\sigma$ on flat spacetime and purity in de Sitter spacetime can also be computed similarly.

To compute $\gamma_\phi$ on flat spacetime for $m_\phi=m_\sigma=0$, we need to perform the integral,
\begin{align}
\label{eq:app_integral}
    I_\phi(p) = \frac{1}{8f^4 p}\int_{\bsk_1, \bsk_2}\frac{1}{k_1k_2|\bsp + \bsk_1 + \bsk_2|}\left(\frac{pk_1 -\bsp\cdot\bsk_1}{p + k_1 + k_2 + |\bsp + \bsk_1 + \bsk_2|}\right)^2\,,
\end{align}
over the integral region specified by the conditions,
\begin{align}
0 \leq k_1 \leq \LUV\,,
\quad
0 \leq k_2 \leq \LUV\,,
\quad
0 \leq |\bsk_1 + \bsk_2 + \bsp| \leq \LUV\,.
\end{align}
For this, it is convenient to parameterize $\bsk_1$ by its magnitude $k_1$, the angle $\theta_1$ with respect to $\bsp$, and the azimuthal angle $\phi_1$ around $\bsp$. On the other hand, to parameterize $\bsk_2$, we use the intermediate vector $\bsl \coloneqq \bsp + \bsk_1$ as a reference vector. Similarly to the $\bsk_1$ case, we introduce the angle $\theta_2$ with respect to $\bsl$ and the azimuthal angle $\phi_2$ around $\bsl$. See Fig.~\ref{fig:momentum_config}. Then, the current setup allows us to perform the following change of variables that exploits the rotational symmetry of the integrand:
\begin{align}
\nonumber
\int \frac{\dd^3\bsk_1}{(2\pi)^3} \, \frac{\dd^3 \bsk_2}{(2\pi)^3} &= \frac{1}{(2\pi)^6}\int k_1^2 \dd k_1 \int \dd \theta_1 \sin\theta_1 \int \dd \phi_1 \int k_2^2 \dd k_2 \int \dd \theta_2 \sin\theta_2 \int \dd\phi_2\\
&=\frac{1}{(2\pi)^4} \int k_1^2 \dd k_1 \int \dd (\cos\theta_1) \int k_2^2 \dd k_2 \int\dd(\cos\theta_2)\,,
\end{align}
where at the second equality we used the fact that the integrand of~\eqref{eq:app_integral} is independent of $\phi_1$ and $\phi_2$.

\begin{figure}[t]
\centering
\resizebox{0.9\textwidth}{!}{
\begin{tabular}{cc}
\begin{tikzpicture}
    \begin{scope}
    \coordinate (O) at (0,0);
    \coordinate (A) at (-3,2);   
    \coordinate (B) at (0,3);    
    \coordinate (C) at (3,1);     
    \coordinate (exA) at (-3.75,2.5);
    \coordinate (exB) at (0,4);
    
    \draw[-{Stealth}, blue, thick] (O) -- node[label,font=\small, below=2pt] {$\bsp$} (A);
    \draw[-{Stealth}, red, thick] (A) -- node[label,font=\small, above] {$\bsk_1$} (B);
    \draw[-{Stealth}, green!60!black, thick] (B) -- node[label,font=\small, above] {$\bsk_2$} (C);
    \draw[-{Stealth}, magenta, thick] (C) -- node[label,font=\small, below] {$\bsk_3$} (O);
        
    \draw[-{Stealth}, orange, dashed, thick] (O) -- node[label,font=\small, right] {$\vec{l}$} (B);
        
    \draw[gray, dashed] (A) -- (exA);
    \draw[gray, dashed] (B) -- (exB);

    \draw[white] (O) -- (3.75,-1.5);

    \draw pic[draw=red, red, angle radius=3mm, angle eccentricity=1.8, "$\theta_1$", font=\small] {angle = B--A--exA};
    \draw pic[draw=green!60!black, green!60!black, angle radius=3mm, angle eccentricity=1.8, "$\theta_2$", font=\small] {angle = C--B--exB};
    \end{scope}
\end{tikzpicture}
&
\tdplotsetmaincoords{60}{130} 
\begin{tikzpicture}[tdplot_main_coords, scale=2.0,
vector/.style={-{Stealth[scale=0.9]}, thick},
    proj/.style={dashed, gray!50, line width=0.4pt},
    label/.style={font=\scriptsize, inner sep=0.5pt},
    extline/.style={dashed, gray!50, line width=0.4pt}]

    \coordinate (O) at (0,0,0);
    \coordinate (A) at (3,0,0);      
    \coordinate (B) at (4.5,2.0,3.5);    
    \coordinate (C) at (1.0,0.5,0);     
    \coordinate (D) at (4.0,0,0);
    
    \draw[vector, blue] (O) -- 
    node[red, label, font=\small, left=15pt] {$\phi_1$}
    node[label, font=\small, below=2pt] {$\bsp$} (A) 
    node[red, pos=0.7] {\AxisRotator[rotate=25,x=0.15cm,y=0.3cm,solid,thin]};
    \draw[vector, red] (A) -- node[label, font=\small, above left=2pt] {$\bsk_1$} (B);
    \draw[vector, green!60!black] (B) -- node[label, font=\small, below left] {$\bsk_2$} (C);
    \draw[vector, magenta] (C) -- node[label, font=\small, right=2pt] {$\bsk_3$} (O);
    
    \begin{scope}[proj]
    \coordinate (Bxy) at ($(B|-O)$);
    \draw (B) -- (Bxy);
    \coordinate (Bxz) at ($(B-|O)$);
    \draw (B) -- (Bxz);
    \coordinate (Cxproj) at (1.0,0,0);
    \draw (C) -- (Cxproj);
    \coordinate (Cyproj) at (0,0.5,0);
    \draw (C) -- (Cyproj);
    \coordinate (Czproj) at (0,0,0);
    \draw (C) -- (Czproj);
    \draw (O) -- (Bxy) -- (A);
    \end{scope}
    
    \draw[orange, dashed, vector] (O) -- 
    node[label, font=\small, above right=2pt] {$\vec{l}$} (B)
    node[green!60!black,label, font=\small, right=11pt] {$\phi_2$}
    node[green!60!black, pos=0.7] {\AxisRotator[rotate=155,x=0.15cm,y=0.3cm,solid,thin]} ;
    
    \coordinate (Lext) at ($(O)!1.5!(B)$);
    \draw[extline] (B) -- (Lext);
    
    \pic[draw=red, angle radius=3mm, angle eccentricity=2, "$\theta_1$"{red}, font=\small] {angle = B--A--D};
    \pic[draw=green!60!black, "$\theta_2$"{shift = (-100:0.20),green!60!black}, angle radius=3mm, angle eccentricity=1.7, font=\small] {angle = Lext--B--C};
    
    \begin{scope}[gray!60, -{Stealth[scale=0.7]}, line width=0.5pt]
    \draw (0,0,0) -- (3.5,0,0) node[below right] {$x$};
    \draw (0,0,0) -- (0,1.5,0) node[right] {$y$};
    \draw (0,0,0) -- (0,0,1.5) node[above] {$z$};
    \end{scope}
\end{tikzpicture}
\end{tabular}
}
\caption{
Momentum configuration compatible with $\bsp + \bsk_1 + \bsk_2 + \bsk_3 = 0$: The left figure illustrates a planar configuration, whereas the right figure is for generic $\phi_2$.
}
\label{fig:momentum_config}
\end{figure}

In order to make the condition $0\leq |\bsk_1 + \bsk_2 + \bsp| \leq \LUV$ more transparent, we further change the integration variables from $\theta_1$, $\theta_2$ to $l \coloneqq |\bsl|$ and $k_3$ by using 
\begin{align}
l= \sqrt{p^2 + k_1^2 + 2pk_1\cos\theta_1}
    \quad &\Rightarrow \quad \dd (\cos\theta_1) = \frac{l}{pk_1}\dd l \,,\\
k_3 = \sqrt{l^2 + k_2^2 + 2lk_2\cos\theta_2} \quad
    &\Rightarrow \quad \dd (\cos\theta_2) = \frac{k_3}{lk_2} \dd k_3
\end{align}
and then we impose the UV cutoff on $p$, $k_1$, $k_2$, $k_3$, respectively. Now we have
\begin{align}
I_\phi(p) 
& = \frac{1}{8(2\pi)^4f^4} 
\int \dd k_1 \int \dd (\cos\theta_1) \int \dd k_2 \int\dd(\cos\theta_2) 
\frac{k_1 k_2}{p k_3}\left(\frac{pk_1 -pk_1\cos\theta_1}{p + k_1 + k_2 + k_3}\right)^2 \notag\\
& = \frac{1}{32(2\pi)^4 f^4 p^2
}\int_0^{\LUV} \dd k_1 \int_{|p-k_1|}^{p+k_1} \dd l 
\int \dd k_2 \int \dd k_3\frac{\left[(p+k_1)^2-l^2 \right]^2}{(p+k_1+k_2+k_3)^2}\,,
\end{align}
where the integral region of $l$ is specified by the triangle inequality. Also, the integral region of $k_2$ and $k_3$ is the domain that satisfies all the following conditions:
\begin{align}
|l-k_2|\leq k_3\leq l+k_2
\,,
\quad
0\leq k_2\leq\LUV\,,
\quad
0\leq k_3\leq\LUV\,.
\end{align}
Note that, in contrast to $\bsp,\bsk_1,\bsk_2$, and $\bsk_3$, the intermediate momentum $\bsl$ is not directly bounded by the UV cutoff, because it is not a momentum carried by the physical modes. Therefore, its integral region is not limited to $l\leq\LUV$, but rather includes $l>\LUV$ as well. In particular, the integral region of $k_2,k_3$ for the latter case is qualitatively different from the former case. See Fig.~\ref{fig:integration}.

\begin{figure}[t]
\centering
\resizebox{0.8\textwidth}{!}{
\begin{tabular}{cc}
\begin{tikzpicture}
    \begin{scope}[shift={(0,0)}]
        \draw[thick,->] (0,0) -- (6,0) node[right] {$k_1$};
        \draw[thick,->] (0,0) -- (0,6) node[above] {$l$};
        
        \node[left] at (0,0) {0};
        \node[left] at (0,2) {$p$};
        \node[left] at (0,4) {$\LUV$};
        \node[below] at (2,0) {$p$};
        \node[below] at (4,0) {$\LUV$};
        
        \draw[dashed] (4,0) -- (4,6);
        \draw[dashed] (0,4) -- (6,4);
        
        \draw[red,thick] (0,2) -- (4,6);
        \draw[red,thick] (0,2) -- (2,0);
        \draw[red,thick] (2,0) -- (6,4);
        
        \fill[green!20,opacity=0.5] (2,4) -- (4,4) -- (4,2) -- (2,0) -- (0,2) -- cycle;
        
        \node[red,above] at (1,4) {$l = p + k_1$};
        \node[red,below] at (5.1,1) {$l = |p - k_1|$};
    \end{scope}
\end{tikzpicture}
&
\begin{tikzpicture}
    \begin{scope}[shift={(8,0)}]
        \draw[thick,->] (0,0) -- (6,0) node[right] {$k_2$};
        \draw[thick,->] (0,0) -- (0,6) node[above] {$k_3$};
        
        \node[left] at (0,0) {0};
        \node[left] at (0,3) {$l$};
        \node[left] at (0,4) {$\LUV$};
        \node[below] at (3,0) {$l$};
        \node[below] at (4,0) {$\LUV$};
        
        \draw[dashed] (4,0) -- (4,6);
        \draw[dashed] (0,4) -- (6,4);
        
        \draw[red,thick] (0,3) -- (3,6);
        \draw[red,thick] (0,3) -- (3,0);
        \draw[red,thick] (3,0) -- (6,3);
        
        \fill[blue!20,opacity=0.5] (1,4) -- (4,4) -- (4,1) -- (3,0) -- (0,3) -- cycle;
        
        \node[red,above] at (1.1,5) {$k_3 = l + k_2$};
        \node[red,below] at (5.2,1) {$k_3 = |l - k_2|$};
    \end{scope}
\end{tikzpicture}
\\
\noalign{
\vskip 1ex
\centerline{\textbf{Case 1}: $l\leq\LUV$}
\vskip 2ex
}
\begin{tikzpicture}
    \begin{scope}[shift={(0,0)}]
        \draw[thick,->] (0,0) -- (6,0) node[right] {$k_1$};
        \draw[thick,->] (0,0) -- (0,6) node[above] {$l$};
        
        \node[left] at (0,0) {0};
        \node[left] at (0,2) {$p$};
        \node[left] at (0,4) {$\LUV$};
        \node[below] at (2,0) {$p$};
        \node[below] at (4,0) {$\LUV$};
        
        \draw[dashed] (4,0) -- (4,6);
        \draw[dashed] (0,4) -- (6,4);
        
        \draw[red,thick] (0,2) -- (4,6);
        \draw[red,thick] (0,2) -- (2,0);
        \draw[red,thick] (2,0) -- (6,4);
        
        \fill[green!20,opacity=0.5] (2,4) -- (4,4) -- (4,6) -- cycle;
        
        \node[red,above] at (1,4) {$l = p + k_1$};
        \node[red,below] at (5.1,2) {$l = |p - k_1|$};
    \end{scope}
\end{tikzpicture}
&
\begin{tikzpicture}
    \begin{scope}[shift={(8,0)}]
        \draw[thick,->] (0,0) -- (6,0) node[right] {$k_2$};
        \draw[thick,->] (0,0) -- (0,6) node[above] {$k_3$};
        
        \node[left] at (0,0) {0};
        \node[left] at (0,4) {$\LUV$};
        \node[left] at (0,5) {$l$};
        \node[below] at (4,0) {$\LUV$};
        \node[below] at (5,0) {$l$};
        
        \draw[dashed] (4,0) -- (4,6);
        \draw[dashed] (0,4) -- (6,4);
        
        \draw[red,thick] (0,5) -- (1,6);
        \draw[red,thick] (0,5) -- (5,0);
        \draw[red,thick] (5,0) -- (6,1);
        
        \fill[blue!20,opacity=0.5] (1,4) -- (4,4) -- (4,1) -- cycle;
        
        \node[red,above] at (1.5,4.5) {$k_3 = l + k_2$};
        \node[red,below] at (5.2,2) {$k_3 = |l - k_2|$};
    \end{scope}
\end{tikzpicture}
\\
\noalign{
\vskip 1ex
\centerline{\textbf{Case 2}: $l>\LUV$}
\vskip 1ex
}
\end{tabular}
}
\caption{Integral regions for $l\leq\LUV$ and $l>\LUV$.}
\label{fig:integration}
\end{figure}

Finally, we make the integration variables dimensionless by dividing them by $\LUV$ as $\pL\coloneqq p/\LUV$, $k_{1\Lambda}\coloneqq k_1/\LUV$, $k_{2\Lambda}\coloneqq k_2/\LUV$, $l_{\Lambda}\coloneqq l/\LUV$, and  $k_{3\Lambda}\coloneqq k_3/\LUV$. Then, we can factor out an $f$-independent dimensionless function $F_\phi^\text{flat}(\pL)$ as
\begin{align}
I_\phi(p) =\left(\frac{\LUV}{f}\right)^4 F^{\text{flat}}_\phi(\pL)
\end{align}
with $F_\phi^\text{flat}(\pL)$ given by
\begin{align}
\label{F_flat_def}
F^{\text{flat}}_\phi(\pL) = 
\frac{1}{32(2\pi)^4 \pL^2}
\int_0^1 \dd k_{1\Lambda}
\int_{|\pL-k_{1\Lambda}|}^{\pL+k_{1\Lambda}} \dd l_\Lambda
\int \dd k_{2\Lambda}
\int \dd k_{3\Lambda}
\frac{\left[(\pL+k_{1\Lambda})^2-l_\Lambda^2\right]^2}{(\pL+k_{1\Lambda}+k_{2\Lambda}+k_{3\Lambda})^2}\,.
\end{align}
Here, the integration region of $k_{2\Lambda}, k_{3\Lambda}$ is specified by
\begin{align}
|\,l_\Lambda-k_{2\Lambda}|\leq k_{3\Lambda}\leq l_\Lambda+k_{2\Lambda}
\,,
\quad
0\leq k_{2\Lambda}\leq1\,,
\quad
0\leq k_{3\Lambda}\leq1\,.
\end{align}
The integral~\eqref{F_flat_def} can be performed analytically, and the result is given in \eqref{C:phi_flat}.

\newpage

\begin{figure}[t]
\centering
  \begin{minipage}[b]{0.49\linewidth}
    \centering
    \includegraphics[width=7cm]{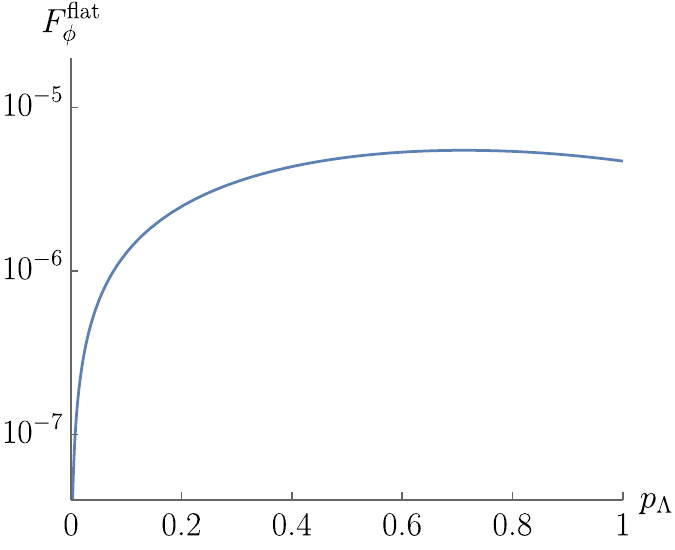}
  \end{minipage}
  \begin{minipage}[b]{0.49\linewidth}
    \centering
    \includegraphics[width=7cm]{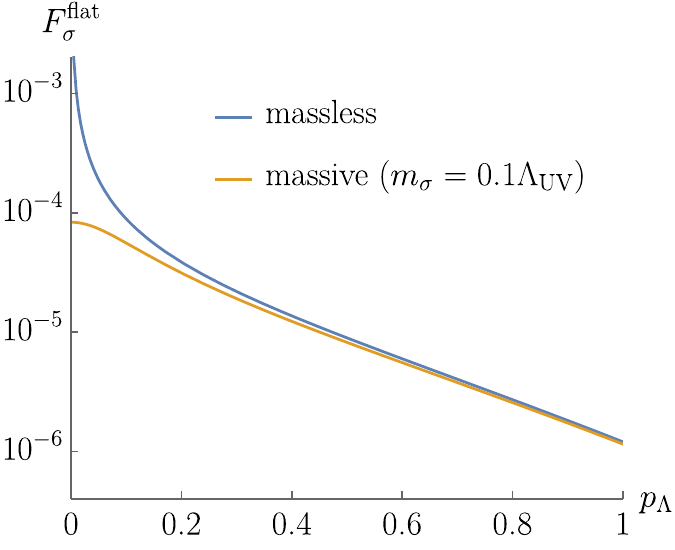}
  \end{minipage}
  \caption{
  Profiles of $F^{\text{flat}}_\phi$ (left) and $F^{\text{flat}}_\sigma$ (right). In the right panel, $F^{\text{flat}}_\sigma$ for the massive $\sigma$ case with $m_\sigma=0.1\,\LUV$ is also plotted numerically (orange curve).
  }
  \label{fig:Fp_flat}
\end{figure}

\section{Explicit form of purity}\label{ExpPurF}

\paragraph{Purity in flat spacetime.}

An explicit form of $\gamma_{\phi}$ in flat spacetime for $m_\phi=m_\sigma=0$ is
\begin{align}
    \gamma_{\phi}(p) = 1 - \left(\frac{\LUV}{f}\right)^4 F^{\text{flat}}_\phi(\pL)
\end{align}
with
\begin{align}\label{C:phi_flat}
    F^{\text{flat}}_\phi(\pL)
    & = \frac{1}{(2\pi)^4}\bigg[ -\frac{3283 \pL^4}{43200}-\frac{791 \pL^3}{1800}-\frac{91 \pL^2}{180}+\frac{28 \pL}{135} -\frac{1}{6} -\frac{22}{45 \pL} \notag\\
    &
    \hspace{48pt}
    - \frac{\pL}{9} \log\left(\pL+1\right) +\left(\frac{p}{9}+\frac{2}{3}+\frac{16}{15 \pL}+\frac{44}{45 \pL^2}\right) \log\left(\frac{2(\pL+1)}{\pL+2}\right) \notag\\
    &
    \hspace{48pt}
    +\left(\frac{11\pL^4}{180}+\frac{13 \pL^3}{30}+\frac{13 \pL^2}{12}+\pL\right) \log\left(\frac{4(\pL+1)^2}{(\pL+2)(\pL+3)}\right)\bigg]\,.
\end{align}
On the other hand, $\gamma_{\sigma}$ in the same setup is given by
\begin{align}
    \gamma_{\sigma}(p) = 1 - \left(\frac{\LUV}{f}\right)^4 F^{\text{flat}}_\sigma(\pL)
\end{align}
with
\begin{align}\label{C:sig_flat}
    F^{\text{flat}}_\sigma(\pL)
    & =\frac{1}{(2\pi)^4}\bigg[ -\frac{979 \pL^4}{9600}-\frac{2971 \pL^3}{4800}-\frac{397 \pL^2}{360}-\frac{47 \pL}{160}+\frac{61}{576}-\frac{241}{960 \pL} \\
    &
    \hspace{48pt}
    -\left(\frac{\pL}{6}+\frac{5}{24}\right) \log (\pL+1)
    \notag\\
    &
    \hspace{48pt}
    +\left(\frac{\pL}{6}+\frac{7}{8}+\frac{16}{15 \pL}+\frac{8}{15 \pL^2}\right) \log \left(\frac{2 (\pL+1)}{\pL+2}\right)\notag\\
    &
    \hspace{48pt}
    +\left(\frac{3 \pL^4}{40}+\frac{3 \pL^3}{5}+\frac{11 \pL^2}{6}+ \frac{5 \pL}{2}+\frac{9}{8}\right) \log \left(\frac{4 (\pL+1)^2}{(\pL+2) (\pL+3)}\right)\bigg]\,. \notag
\end{align}
See Fig.~\ref{fig:Fp_flat} for profiles of $F^{\text{flat}}_\phi(\pL)$ and $F^{\text{flat}}_\sigma(\pL)$. There, we also plotted a numerical result of $F^{\text{flat}}_\sigma(\pL)$ for $m_\phi=0$ and $m_\sigma=0.1\,\LUV$, which demonstrates that $F^{\text{flat}}_\sigma(\pL)$ becomes regular at $\pL = 0$ by giving a mass to $\sigma$.

\newpage

\paragraph{Purity in de Sitter spacetime.}

In de Sitter spacetime, $\gamma_{\phi}(p)$ for $m_\phi=0$, $m_\sigma=\sqrt{2}H$ is given by
\begin{align}
    \gamma_\phi(p) = 1 - \left[\left(\frac{\LUV}{f}\right)^4 F_{\phi0}^{\text{dS}}\left(\bpL \right) + \left(\frac{H}{f}\right)^2 \left(\frac{\LUV}{f}\right)^2 F_{\phi2}^{\text{dS}} \left(\bpL \right) +\left(\frac{H}{f}\right)^4 F_{\phi4}^{\text{dS}}\left(\bpL \right) \right]\,,
\end{align}
with
\begin{align}
    F^{\text{dS}}_{\phi0}(\bpL)
    & = F^{\text{flat}}_\phi(\pL) \Big|_{\pL = \bpL}\,,\label{C:phi0_dS}\\
    F^{\text{dS}}_{\phi2}(\bpL)
    & = \frac{1}{(2\pi)^4}\bigg[\frac{383 \bpL^3}{7200}-\frac{407 \bpL^2}{8640}+\frac{11 \bpL}{450}+\frac{307}{720}+\frac{923}{1080 \bpL}+\frac{11}{30 \bpL^2}-\frac{22}{45 \bpL^3} \notag\\
    &
    \hspace{48pt}
    -\frac{1}{36 \bpL}\log (\bpL+1)-\left(\frac{2}{3}+\frac{89}{36 \bpL}+\frac{31}{15 \bpL^2}-\frac{44}{45 \bpL^4}\right) \log \left(\frac{2 (\bpL+1)}{\bpL+2}\right) \notag\\*
    &
    \hspace{48pt}
    -\left(\frac{\bpL^3}{30}+\frac{\bpL^2}{18}-\frac{\bpL}{60}-\frac{1}{4\bpL}\right) \log \left(\frac{4 (\bpL+1)^2}{(\bpL+2) (\bpL+3)}\right)\bigg]
    \,,\label{C:phi2_dS}\\
    F^{\text{dS}}_{\phi4}(\bpL)
    & =\frac{1}{(2\pi)^4}\bigg[ \frac{1}{(\bpL+1)^3 (\bpL+3)^3} \bigg(\frac{101 \bpL^7}{2400}+\frac{797 \bpL^6}{1440}+\frac{11381 \bpL^5}{5400}-\frac{1951 \bpL^4}{960}-\frac{266633 \bpL^3}{7200}
    \notag\\
    &
    \hspace{48pt}
    -\frac{4845461 \bpL^2}{43200}-\frac{1126849 \bpL}{7200}-\frac{32479}{320}-\frac{86}{5 \bpL}+\frac{3021}{320 \bpL^2}+\frac{459}{160 \bpL^3}\bigg) \notag\\
    &
    \hspace{48pt}
    +\left(\frac{\bpL}{24}-\frac{17}{72 \bpL}-\frac{1}{4 \bpL^2}-\frac{5}{288 \bpL^4}\right) \log (\bpL+1) \notag\\
    &
    \hspace{48pt}
    -\left(\frac{\bpL}{12}+\frac{49}{240}-\frac{157}{180 \bpL}-\frac{31}{10 \bpL^2}+\frac{8}{45 \bpL^4}\right) \log \left(\frac{2 (\bpL+1)}{\bpL+2}\right) \notag\\
    &
    \hspace{48pt}
    -\left(\frac{\bpL}{30}-\frac{17}{36 \bpL}\right) \log \left(\frac{4 (\bpL+1)^2}{(\bpL+2) (\bpL+3)}\right)  \bigg]  
    \,.\label{C:phi4_dS}
\end{align}
\begin{figure}[t]
\centering
  \begin{minipage}[b]{0.49\linewidth}
    \centering
    \includegraphics[width=7cm]{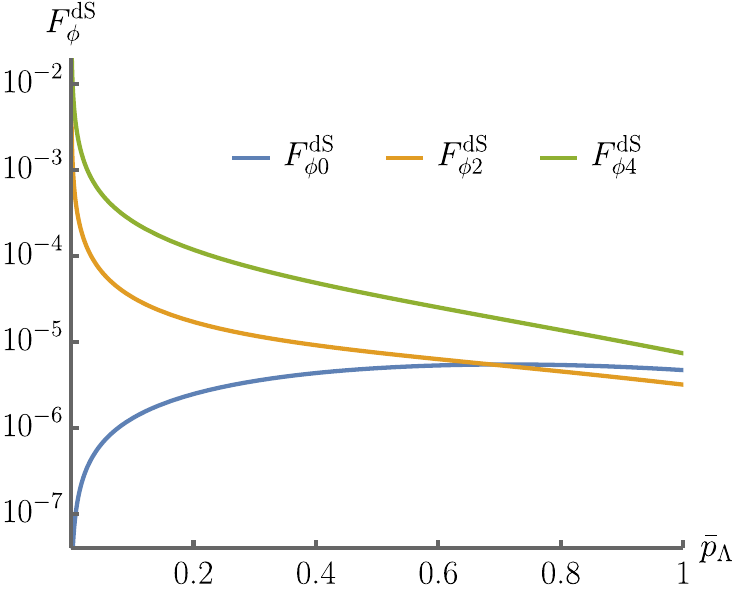}
  \end{minipage}
  \begin{minipage}[b]{0.49\linewidth}
    \centering
    \includegraphics[width=7cm]{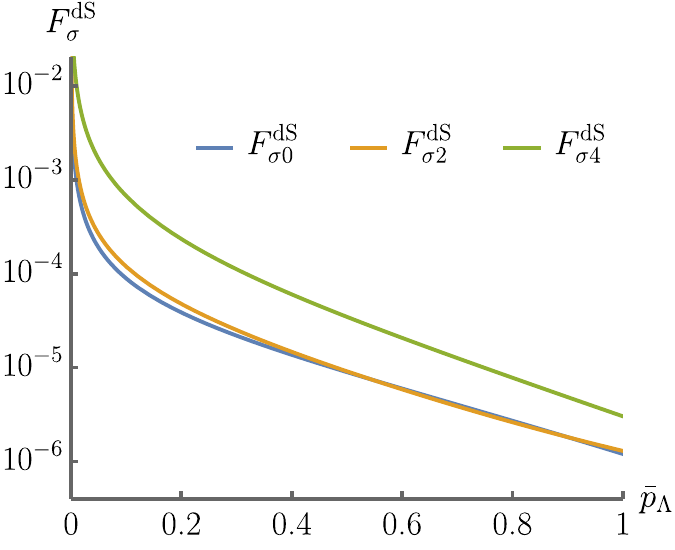}
  \end{minipage}
  \caption{
  Profiles of $F^{\text{dS}}_{\phi0},F^{\text{dS}}_{\phi2},F^{\text{dS}}_{\phi4}$ (left) and $F^{\text{dS}}_{\sigma0},F^{\text{dS}}_{\sigma2},F^{\text{dS}}_{\sigma4}$ (right).
  }
  \label{fig:Fp_dS}
\end{figure}
On the other hand, $\gamma_{\sigma}$ in the same setup is
\begin{align}
    \gamma_\sigma(p) = 1 - \left[\left(\frac{\LUV}{f}\right)^4 F_{\sigma0}^{\text{dS}}\left(\bpL \right) + \left(\frac{H}{f}\right)^2 \left(\frac{\LUV}{f}\right)^2 F_{\sigma2}^{\text{dS}} \left(\bpL \right) +\left(\frac{H}{f}\right)^4 F_{\sigma4}^{\text{dS}}\left(\bpL \right) \right]
\end{align}
with
\begin{align}
    F^{\text{dS}}_{\sigma0}(\bpL)
    & = F^{\text{flat}}_\sigma(\pL) \Big|_{\pL = \bpL}\,,\label{C:sig0_dS}\\
    \label{C:sig2_dS}
    F^{\text{dS}}_{\sigma2}(\bpL)
    & = \frac{1}{(2\pi)^4}\bigg[\frac{671 \bpL^3}{1200}+\frac{575 \bpL^2}{288}+\frac{157 \bpL}{80}-\frac{221}{720}-\frac{61}{160 \bpL} \notag\\
    &
    \hspace{47pt}
    -\left(\frac{5}{12}+\frac{1}{4 \bpL}\right) \log (\bpL+1)-\left(\frac{1}{4}-\frac{1}{12 \bpL}-\frac{1}{5 \bpL^2}\right) \log \left(\frac{2 (\bpL+1)}{\bpL+2}\right) \notag\\
    &
    \hspace{47pt}
    -\left(\frac{2 \bpL^3}{5}+\frac{7 \bpL^2}{3}+\frac{19\bpL}{4}+\frac{7}{2}+\frac{3}{4\bpL}\right) \log \left(\frac{4 (\bpL+1)^2}{(\bpL+2) (\bpL+3)}\right)
    \bigg]
    \,,\\
    F^{\text{dS}}_{\sigma4}(\bpL) 
    & = \frac{1}{(2\pi)^4}\bigg[\frac{1}{(\bpL+1)^3 (\bpL+3)^3}\bigg(-\frac{667 \bpL^8}{960}-\frac{40069 \bpL^7}{4320}-\frac{1470319 \bpL^6}{28800}-\frac{179093 \bpL^5}{1200} \notag\\
    &
    \hspace{48pt}
    -\frac{21278021 \bpL^4}{86400}-\frac{1548331 \bpL^3}{7200}-\frac{317329 \bpL^2}{5760}+\frac{45653 \bpL}{720}+\frac{36601}{640}+\frac{1137}{80 \bpL}\bigg) \notag\\*
    &
    \hspace{48pt}
    + \frac{9}{32} \log (\bpL) +\left(\frac{\bpL}{36}-\frac{5}{32}-\frac{13}{12 \bpL}-\frac{59}{144 \bpL^2}\right) \log (\bpL+1) \notag\\
    &
    \hspace{48pt}
    -\left(\frac{\bpL}{18}-\frac{13}{120}-\frac{17}{30 \bpL}-\frac{263}{180 \bpL^2}\right) \log \left(\frac{2 (\bpL+1)}{\bpL+2}\right) \notag\\
    &
    \hspace{48pt}
    +\left(\frac{\bpL^2}{2}+\frac{61 \bpL}{36}+2+\frac{5}{3 \bpL}\right) \log \left(\frac{4 (\bpL+1)^2}{(\bpL+2) (\bpL+3)}\right)
    \bigg]
    \,.\label{C:sig4_dS}
\end{align}
See Fig.~\ref{fig:Fp_dS} for the profiles of
$F^{\text{dS}}_{\phi0}(\bpL),F^{\text{dS}}_{\phi2}(\bpL),F^{\text{dS}}_{\phi4}(\bpL)$ and
$F^{\text{dS}}_{\sigma0}(\bpL),F^{\text{dS}}_{\sigma2}(\bpL),F^{\text{dS}}_{\sigma4}(\bpL)$.

\newpage

\bibliography{purity}
\bibliographystyle{utphys}

\end{document}